\documentclass[iop]{emulateapj}
\usepackage{amsmath}

\begin{document}
	
\title{Kinetics and mechanisms of the acid-base reaction between NH$_3$ and HCOOH in interstellar ice analogs}

\author{Jennifer B. Bergner}
\affil{Harvard University Department of Chemistry and Chemical Biology}
\affil{10 Oxford Street, Cambridge, MA 02138, USA}
\email{jennifer.bergner@cfa.harvard.edu}
\author{Karin I. \"Oberg}
\affil{Harvard-Smithsonian Center for Astrophysics}
\affil{60 Garden Street, Cambridge, MA 02138, USA}
\author{Mahesh Rajappan}
\affil{Harvard-Smithsonian Center for Astrophysics}
\affil{60 Garden Street, Cambridge, MA 02138, USA}
\author{Edith C. Fayolle}
\affil{Harvard-Smithsonian Center for Astrophysics}
\affil{60 Garden Street, Cambridge, MA 02138, USA}

\def\placetableone{
\begin{deluxetable}{lccc}
\tabletypesize{\footnotesize}
\tablecaption{Summary of experiments}
\tablecolumns{4} 
\tablewidth{0pt} 
\tablehead{\colhead{Expt}                             &
	\colhead{NH$_{3}$ (ML)\tablenotemark{a}}          &
	\colhead{HCOOH (ML)\tablenotemark{a}}             &
	\colhead{Set temp (K)\tablenotemark{b}}           }
\startdata
\sidehead{TPD experiments\tablenotemark{c}} 
1 & 10.0 & -    & N/A \\
2 & -    & 13.2 & N/A \\ 
3 & 9.2  & 6.3  & N/A \\  \hline
\sidehead{Codeposition experiments} 
4 & 17.6 & 6.9 & 20\\ 
5 & 18.5 & 7.5 & 65 \\ 
6 & 18.4 & 7.4 & 85 \\ \hline
\sidehead{Layered step experiments} 
7 & 1.3 & 1.3 & 75-125\tablenotemark{d}	\\
8 & 3.7 & 2.9 & 40-140\tablenotemark{d} \\
9 & 8.6 & 7.5 & 40-140\tablenotemark{d}  \\   \hline 
\sidehead{Layered single-temperature experiments (1ML)}
10 & 1.5 & 1.0 & 75 \\
11 & 1.5 & 1.1 & 85 \\
12 & 1.8 & 1.2 & 95 \\  
13 & 1.8 & 0.9 & 105 \\ \hline
\sidehead{Layered single-temperature experiments (8ML)}
14 & 9.2 & 6.3 & 40 \\ 
15 & 9.3 & 7.9 & 85 \\ 
16 & 9.2 & 6.8 & 95 \\ 
17\tablenotemark{e} & 9.1 & 7.7 & 95 \\  
18 & 9.1 & 7.3 & 105 \\
19 & 9.7 & 6.7 & 115 \\ 
20\tablenotemark{e} & 8.2 & 6.1 & 115 \\
21 & 8.3 & 7.5 & 125 \\ 
22 & 8.4 & 6.8 & 130 \\ 
23 & 9.0 & 6.9 & 140 \\
24 & 9.5 & 6.4 & 115\tablenotemark{f}
\enddata
\tablenotetext{a}{20\% uncertainty due to band strength uncertainty \citep{Bouilloud2015}}
\tablenotetext{b}{$\sim$4hr isothermal hold unless otherwise noted}
\tablenotetext{c}{Continuous heating, 5K/min}
\tablenotetext{d}{Multiple 1hr holds at 10-20K intervals within range}
\tablenotetext{e}{Duplicates for reproducibility}
\tablenotetext{f}{8hr hold}
\label{tab1}
\end{deluxetable}
}
\def\placetabletwo{
\begin{deluxetable*}{lllcc}
\tabletypesize{\footnotesize}
\tablecolumns{4}
\tablewidth{0pt}
\tablecaption{IR band positions and strengths}
\tablehead{\colhead{Molecule} &
	\colhead{IR mode} &
	\colhead{Position} &
	\multicolumn{2}{c}{Band Strength $A_{i}$ (cm molec$^{-1}$)} \\
	\colhead{} &
	\colhead{} &
	\colhead{(cm$^{-1}$)} &
	\colhead{Literature} &
	\colhead{Derived- 15K} }
\startdata
HCOOH & $\nu_{6}$ CO str & 1216 & 2.9 x 10$^{-17}$ \tablenotemark{(a)} & -  \\
NH$_{3}$ & $\nu_{1}$ s-str & 3375 & 3.0 x 10$^{-17}$ \tablenotemark{(a)} & -  \\
HCOO$^{-}$ & $\nu_{2}$ CO a-str & 1573 & 1.3 x 10$^{-16}$ \tablenotemark{(b)} & 1.1 x 10$^{-16}$  \\
& $\nu_{5}$ CH bend & 1377 & - & 1.3 x 10$^{-17}$  \\
& $\nu_{4}$ CO s-str & 1346 & - & 2.1 x 10$^{-17}$ \\
NH$_{4}^{+}$ & $\nu_{4}$ bend & 1476 \tablenotemark{(d)} & 4.4 x 10$^{-17}$ \tablenotemark{(c)} & 4.8 x 10$^{-17}$  
\enddata
\tablenotetext{a}{\citet{Bouilloud2015}}
\tablenotetext{b}{\citet{Galvez2010}}
\tablenotetext{c}{\citet{Schutte2002}}
\tablenotetext{d}{Indicates broad band}
\label{Tab2}
\end{deluxetable*} 
}	

\def\placetablethree{
\begin{deluxetable}{lll}
\tabletypesize{\footnotesize}
\tablecaption{Derived kinetic parameters}
\tablecolumns{3} 
\tablewidth{0pt} 
\tablehead{\colhead{Category}                  &
	\colhead{E$_a$ (K)}                        &
	\colhead{A (s$^{-1}$)}                     }
\startdata
\sidehead{Reaction}
Combined \tablenotemark{*}   & 70 (30) & 1.4 (0.4) x 10$^{-3}$ \\
Thick ice, long hold & 86   & 1.6 x 10$^{-3}$      \\
Thin ice, long hold  & 40   & 9.8 x 10$^{-4}$      \\
Thin ice, short hold  & 95  & 1.7 x 10$^{-3}$      \\
\sidehead{Pre-reaction}
Diffusion-limited  & 770 (110)   & 7.6 (7.6) x 10$^{-3}$ \\
Orientation-limited  & 400 (10)  & 1.5 (0.2) x 10$^{-3}$         		
\enddata
\tablenotetext{*}{Recommended value}
\tablecomments{Uncertainties listed in parentheses where applicable}
\label{Eas}
\end{deluxetable}
}

\def\placetablefour{
\begin{deluxetable*}{lcccc}
\tabletypesize{\footnotesize}
\tablecaption{Kinetic rate constants}
\tablecolumns{5} 
\tablewidth{0pt} 
\tablehead{\colhead{Series}                    &
	\colhead{Expt}                             &
	\colhead{Set temp (K)}                     &
	\colhead{$k_r$ (s$^{-1}$)}                 &
	\colhead{$k_p$ (s$^{-1}$)}                 }
\startdata
Steps (1ML)        & 7  & 75  &  3.7 x 10$^{-4}$  & -               \\
&    & 85  &  4.4 x 10$^{-4}$  & -               \\
&    & 95  &  8.0 x 10$^{-4}$  & -               \\
&    & 105 &  5.5 x 10$^{-4}$  & -               \\
&    & 115 &  3.3 x 10$^{-3}$  & -               \\
&    & 125 &  6.6 x 10$^{-4}$  & -               \\  \hline
Steps (3ML)        & 8  & 40  &  -                & -               \\ 
&   & 60  &  -                & -               \\
&	& 80  &  4.0 x 10$^{-3}$  & 1.3  x 10$^{-5}$ \\
&	& 100 &  -                & -               \\
&	& 120 &  -                & -               \\
&	& 140 &  -                & -               \\  \hline
Steps (8ML)        & 9  & 40  &  -                & -               \\  
&	& 85  &  -                & -               \\  
&	& 105 &  1.9 x 10$^{-3}$  & 1.6 x 10$^{-6}$ \\  
&	& 125 &  -                & -               \\  
&	& 140 &  4.2 x 10$^{-3}$  & 7.1 x 10$^{-5}$ \\  \hline
Single-temp. (1ML) & 10 & 75  &   6.3 x 10$^{-4}$  & 7.4 x 10$^{-6}$ \\
& 11 & 85  &  6.4 x 10$^{-4}$  & 1.4 x 10$^{-5}$ \\
& 12 & 95  &  5.7 x 10$^{-4}$  & 2.3 x 10$^{-5}$ \\  
& 13 & 105 &  8.1 x 10$^{-4}$  & 3.3 x 10$^{-5}$ \\  \hline
Single-temp. (8ML) & 12 & 40  &  -                & -               \\
& 15 & 85  &  5.5 x 10$^{-4}$  & 2.1 x 10$^{-6}$ \\
& 16 & 95  &  9.1 x 10$^{-4}$  & 2.4 x 10$^{-6}$ \\
& 17 & 95  &  7.1 x 10$^{-4}$  & 2.1 x 10$^{-6}$ \\
& 18 & 105 &  5.9 x 10$^{-4}$  & 4.2 x 10$^{-6}$ \\
& 19 & 115 &  1.0 x 10$^{-3}$  & 1.3 x 10$^{-5}$ \\
& 20 & 115 &  7.0 x 10$^{-4}$  & 7.6 x 10$^{-6}$ \\
& 21 & 125 &  7.7 x 10$^{-4}$  & 1.8 x 10$^{-5}$ \\
& 22 & 130 &  9.1 x 10$^{-4}$  & 2.7 x 10$^{-5}$ \\
& 23 & 140 &  1.4 x 10$^{-3}$  & 5.9 x 10$^{-5}$ \\
& 24\tablenotemark{a} & 115 & 7.9 x 10$^{-4}$ & 7.1 x 10$^{-6}$
\enddata
\tablenotetext{a}{8 hr isothermal hold}
\label{ks}
\end{deluxetable*}
}

\def\placefigureexperimental{
	\begin{figure}[h!]
		\centering
		\includegraphics[width= 0.5\textwidth]{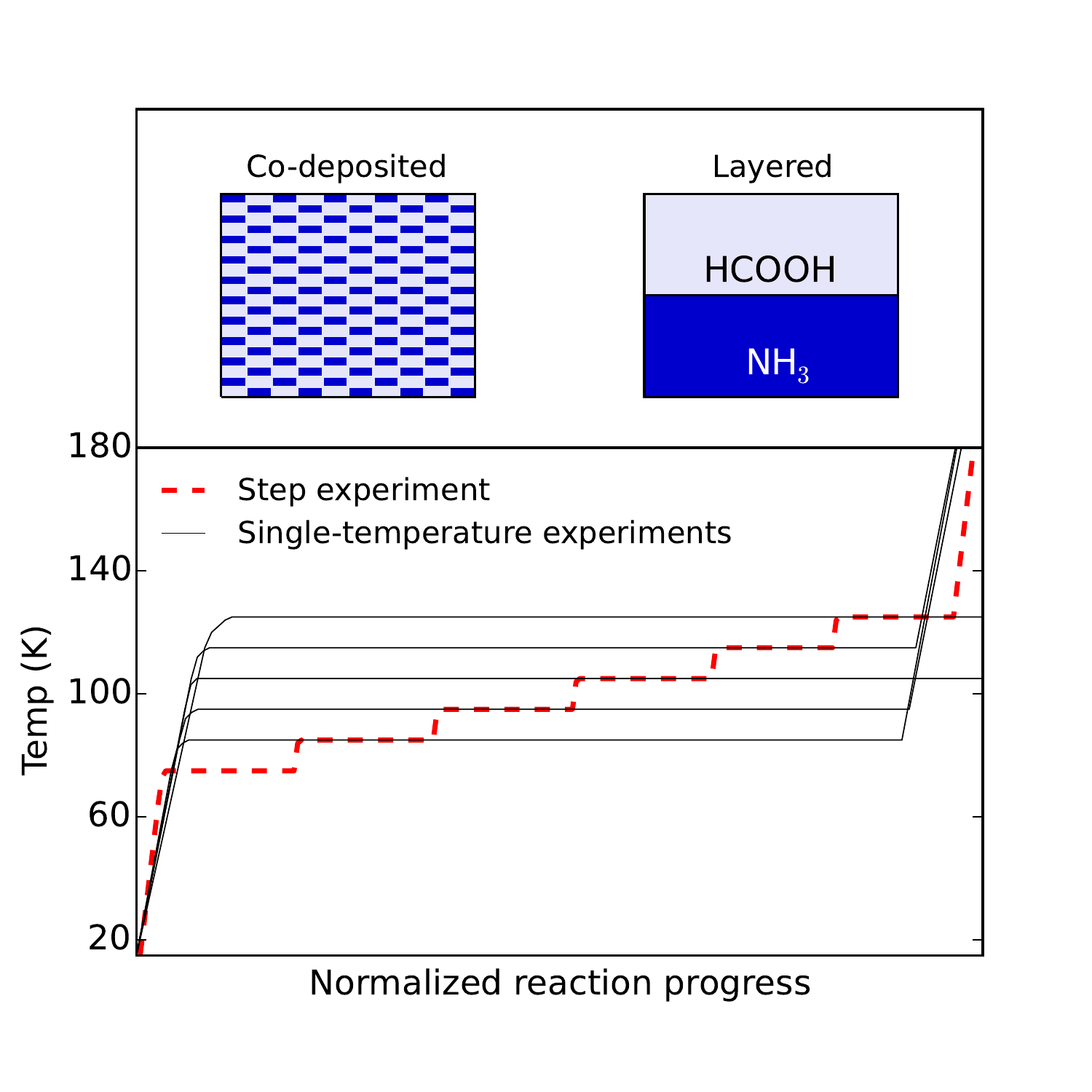}
		\caption{Top panel: schematic of co-deposited vs. layered experimental setups.  NH$_3$ (dark) and HCOOH (light) are either deposited simultaneously to produce an evenly mixed sample, or deposited sequentially to produced a layered sample.  Bottom panel: schematic of step experiments (short isothermal holds) vs. single-temperature experiments (long isothermal holds).}
		\label{experimental}
	\end{figure} 
}
\def\placefigureIR{	
	\begin{figure}[h!]
		\centering
		\includegraphics[width=0.5\textwidth]{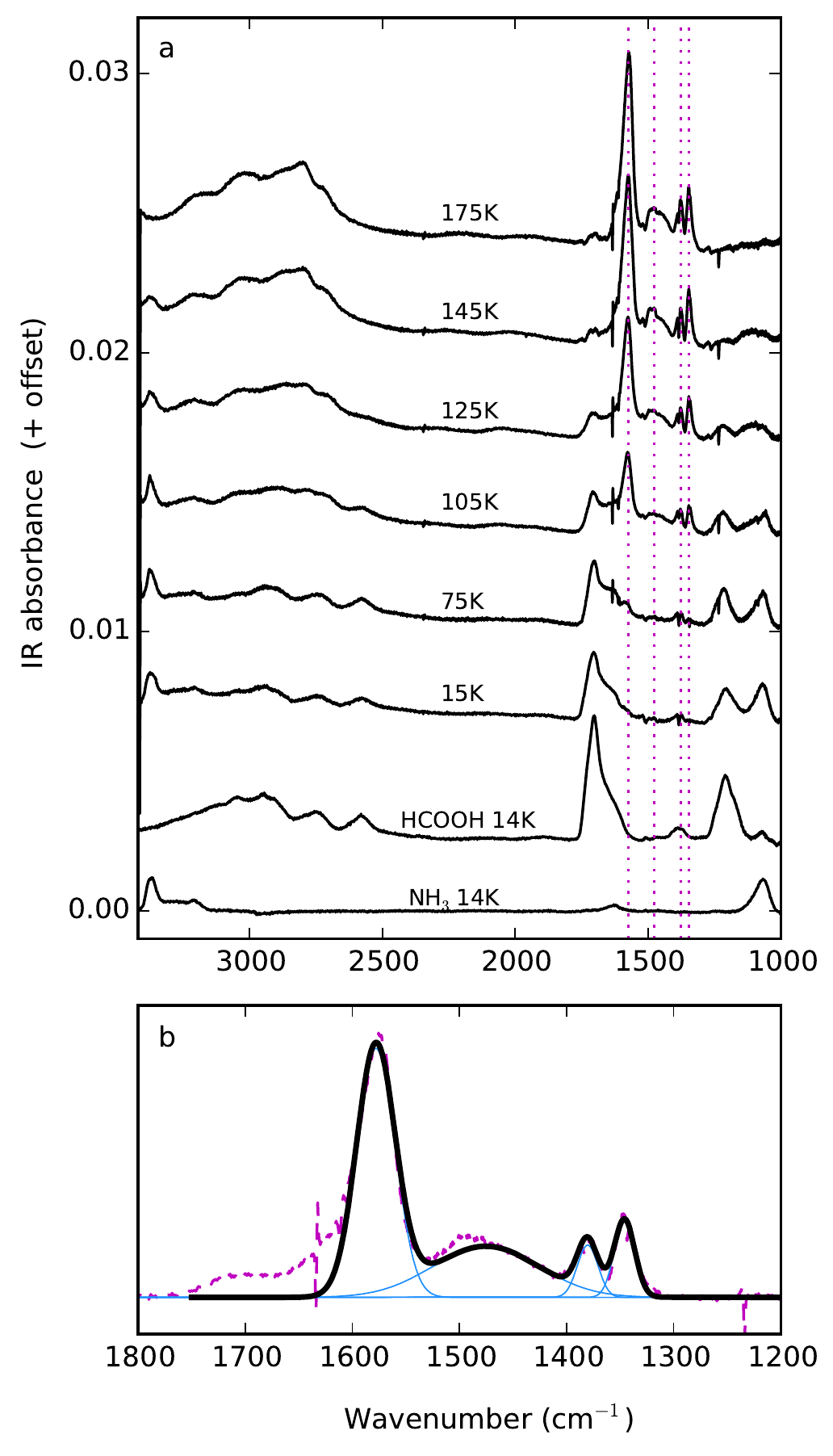}
		\caption{a: IR spectra for pure NH$_{3}$, pure HCOOH, and increasing temperatures of a layered NH$_3$/HCOOH experiment.  Salt features in the 1300-1600 cm$^{-1}$ region (dotted red lines) do not correspond to any feature of the pure ices, and grow as the temperature is increased.  b: zoomed IR spectrum of salt (dashed line) fitted with a 4-component Gaussian.  Solid thin lines represent individual Gaussian fits, and the total fit is shown by the solid thick line.  The outermost peaks exhibit the least overlap.}
		\label{IR}
	\end{figure} 
}
\def\placefigureEdists{
	\begin{figure}[h!]
		\centering
		\includegraphics[width=0.5\textwidth]{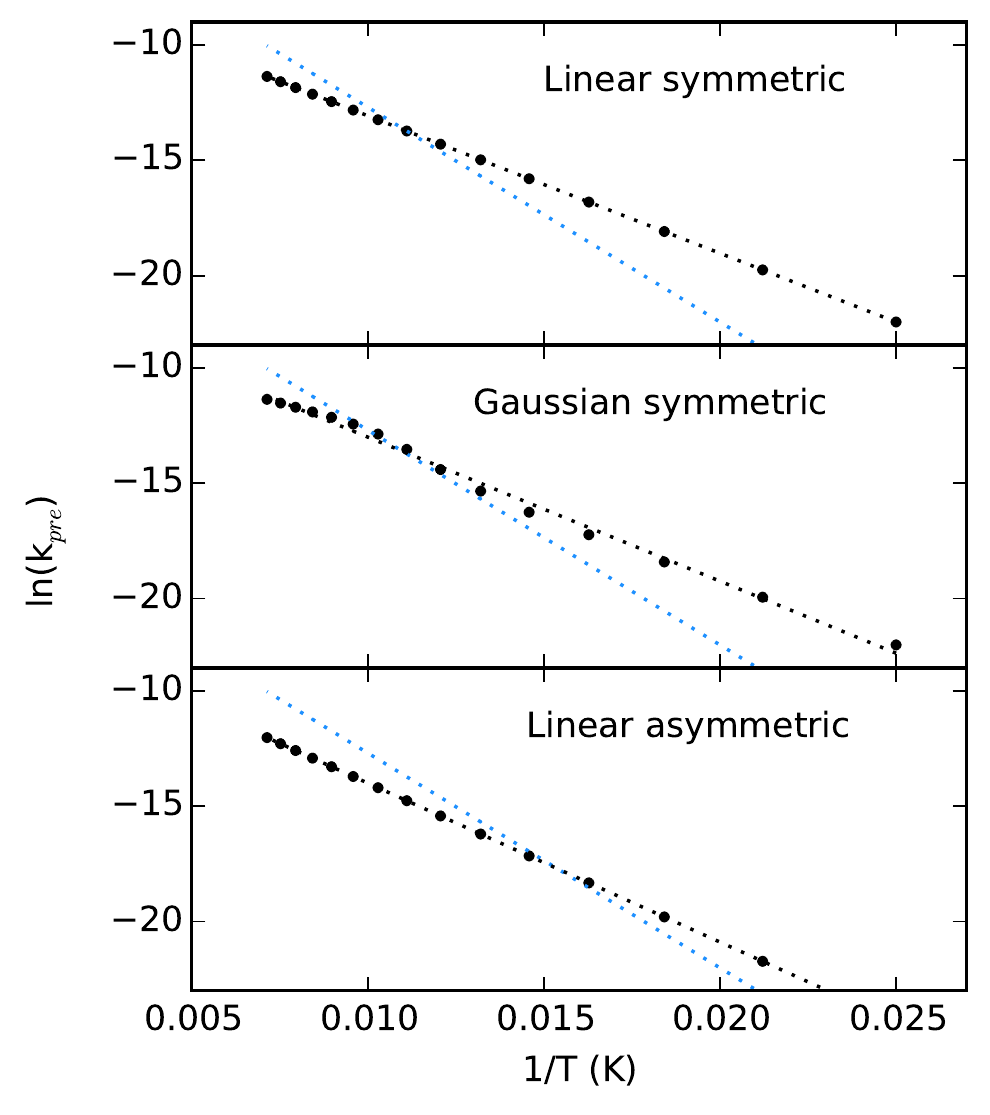}
		\caption{Arrhenius plots for energy barrier distributions.  Rate constants are calculated for a single energy barrier across all temperatures (blue dotted line, all panels) and compared to those calculated using an energy barrier that varies with temperature (circles; best-fits shown as black dotted lines)}
		\label{Edists}
	\end{figure} 
}
\def\placefigureBSexp{
	\begin{figure}[h!]
		\centering
		\includegraphics[width=0.5\textwidth]{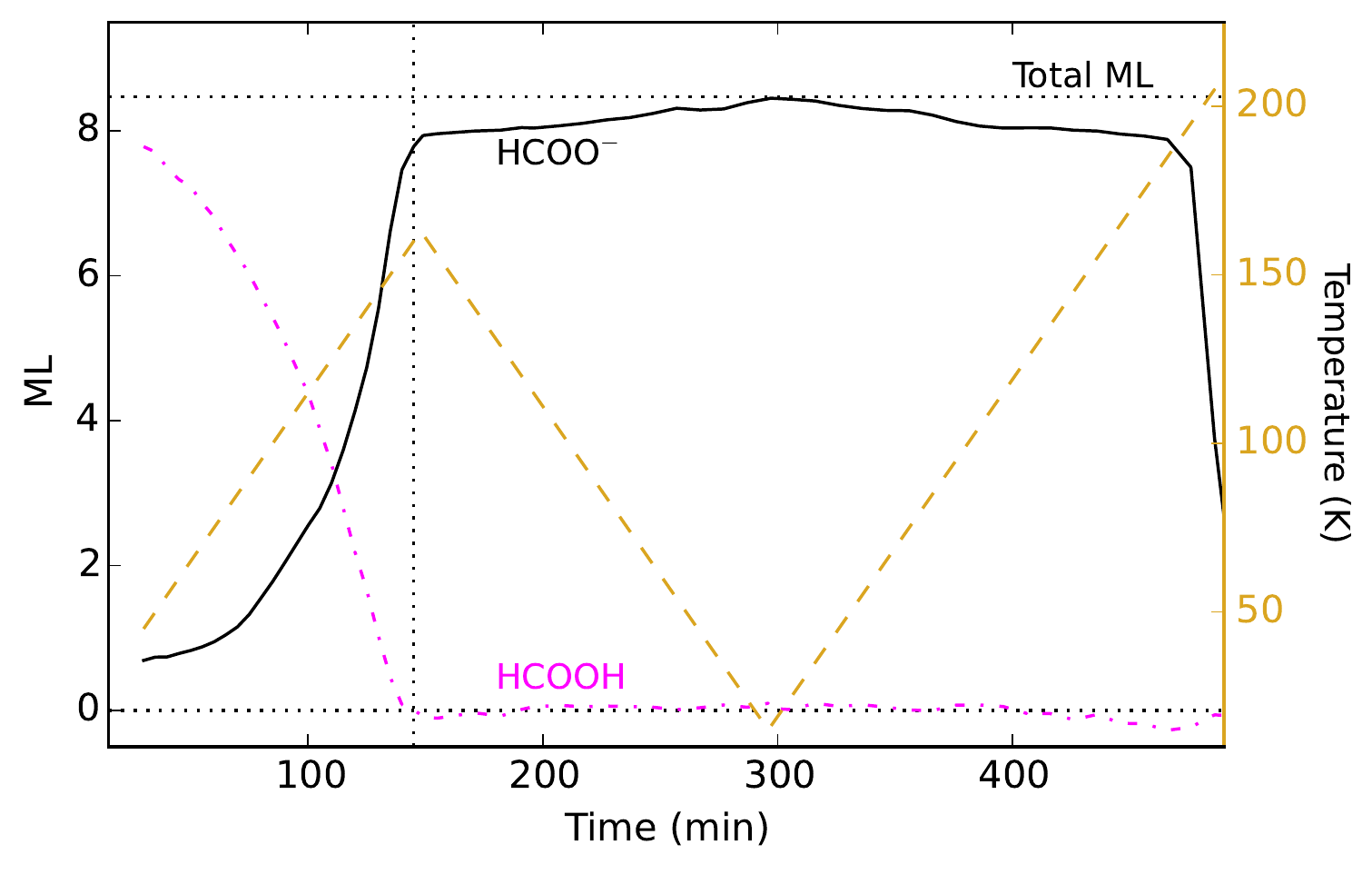}
		\caption{Band strength determination. Salt growth and HCOOH consumption are overplotted with the the temperature profile.  The sample is heated to 165K (vertical dotted line), at which point HCOOH consumption is complete.  The sample is cooled to 14K and re-heated to desorption.  Horizontal dotted lines represent 0ML and the sum total of HCOOH and HCOO$^-$.}
		\label{BSexp}
	\end{figure} 
}
\def\placefigureTPDs{
	\begin{figure}[h!]
		\centering
		\includegraphics[width = 0.5\textwidth]{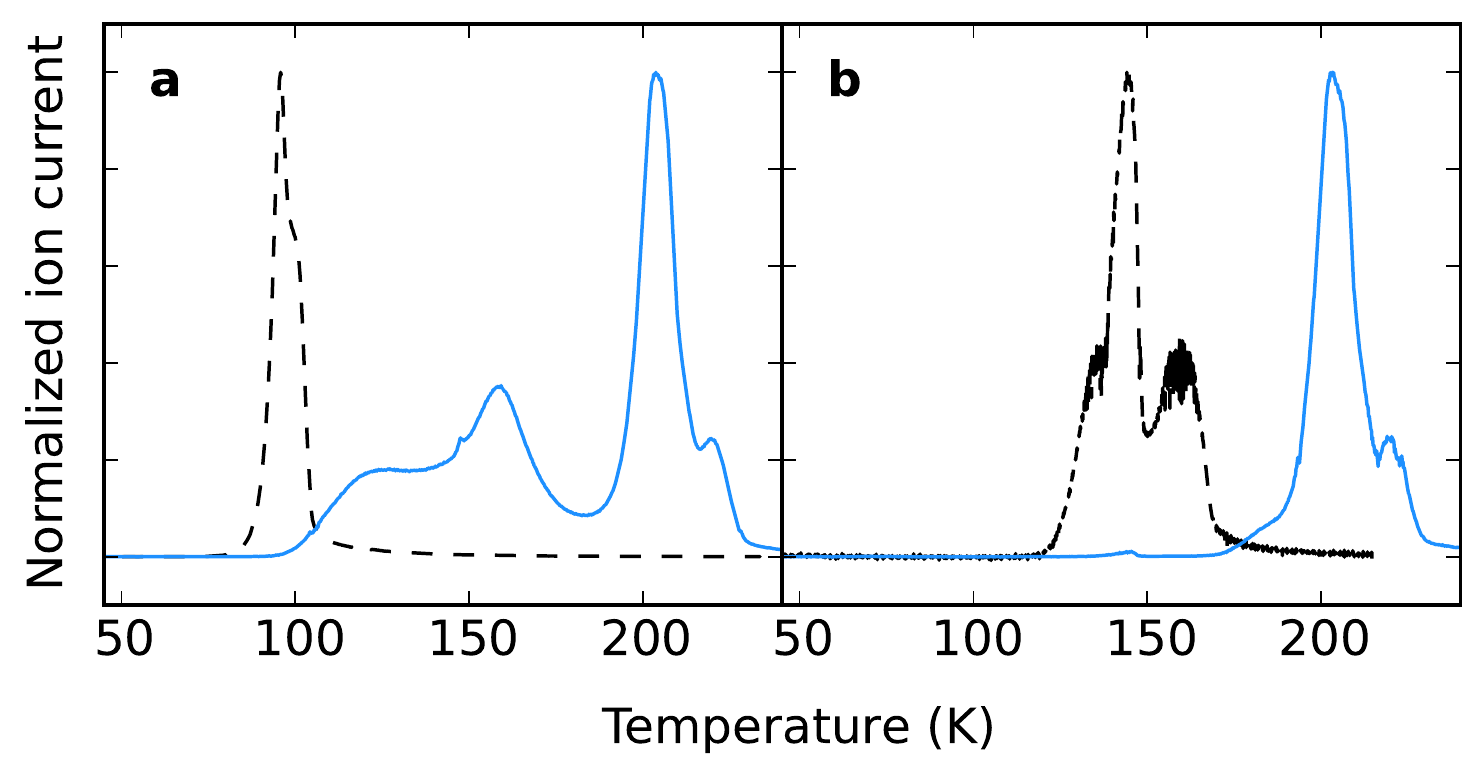}
		\caption{Pure and mixed HCOOH and NH$_3$ TPDs.  Left plot: Mass 17 QMS traces NH$_3$.  Right plot: Mass 29 QMS traces HCOOH.  Dashed lines are for pure ice TPDs, and solid lines are the mixed experiment.}
		\label{TPDs}
	\end{figure} 
}
\def\placefigurecodep{
	\begin{figure}[h!]
		\centering
		\includegraphics[width=0.5\textwidth]{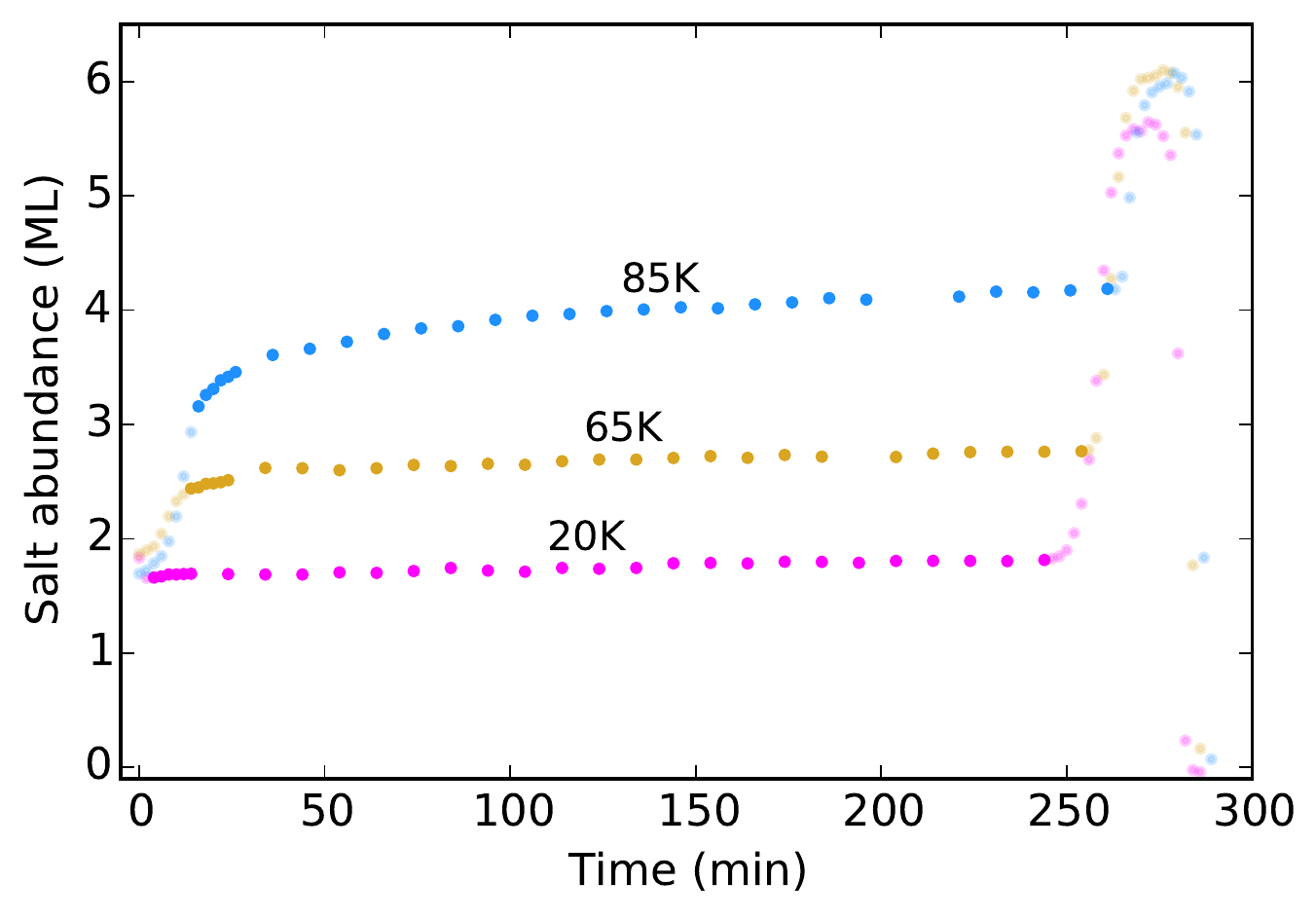}
		\caption{Salt growth in co-deposited ices held at 20K (exp 4; pink circles), 65K (exp 5; gold circles), and 85K (exp 6; blue circles).  Each sample is heated from 14K to the target temperature, held for $\sim$4h at a constant temperature, and heated again until desorption.  Warm-up periods are shown in light colors and the isothermal period in saturated colors for different target temperatures.}
		\label{codep}
	\end{figure} 
}
\def\placefigurecolay{
	\begin{figure}[h!]
		\centering
		\includegraphics[width=0.5\textwidth]{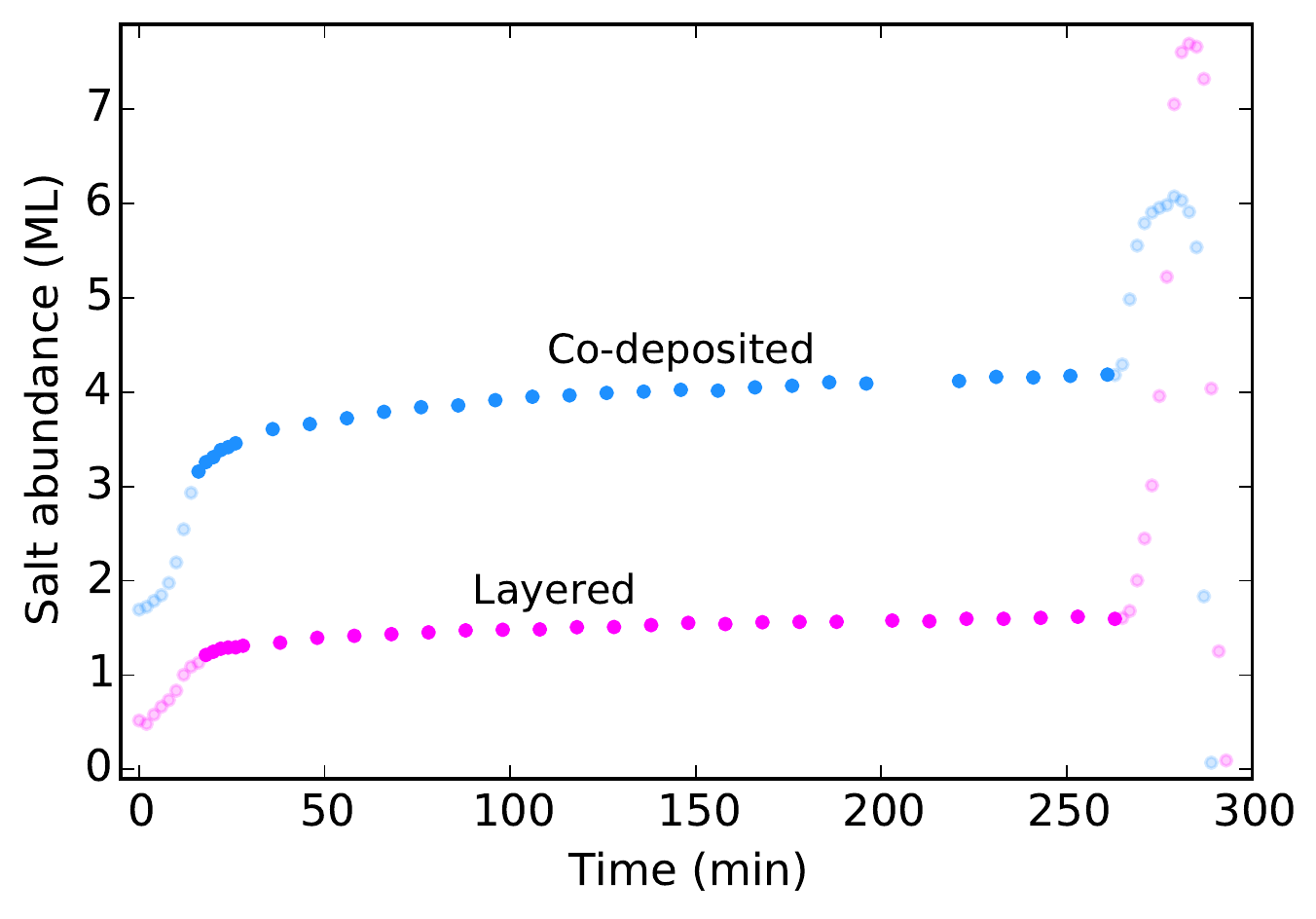}
		\caption{Comparison of salt growth throughout layered (exp 15; pink circles) vs. co-deposited (exp 6; blue circles) experiments, both with an 85K isothermal hold.  Each sample is heated from 14K to 85K, held for $\sim$4h at a constant temperature, and heated again until desorption.  Warm-up periods are shown in light colors and the isothermal period in saturated colors for different target temperatures.}
		\label{colay}
	\end{figure} 
}
\def\placefiguremlstep{
	\begin{figure}[h!]
		\centering
		\includegraphics[width=0.5\textwidth]{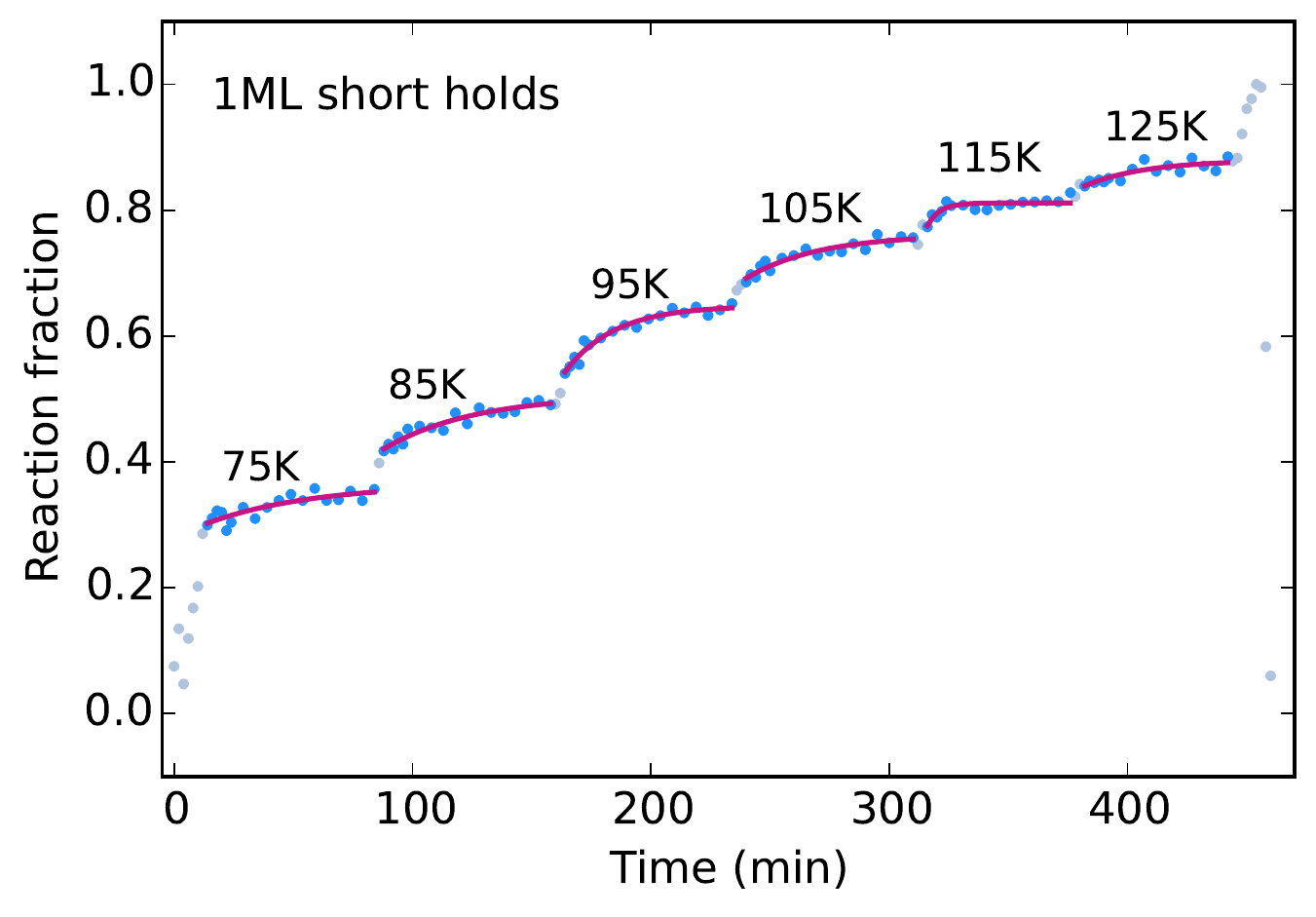}
		\caption{Salt growth throughout the course of the 1ML step experiment (exp 7).  The sample is deposited at 14K and held for $\sim$1h at each target temperature, then is heated to desorption.  Warm-up periods are shown in light colors, the isothermal periods in saturated colors, and the first-order model fits as solid pink lines at each target temperature.}
		\label{1mlstep}
	\end{figure} 
}
\def\placefiguremllong{
	\begin{figure}[h!]
		\centering
		\includegraphics[width=0.5\textwidth]{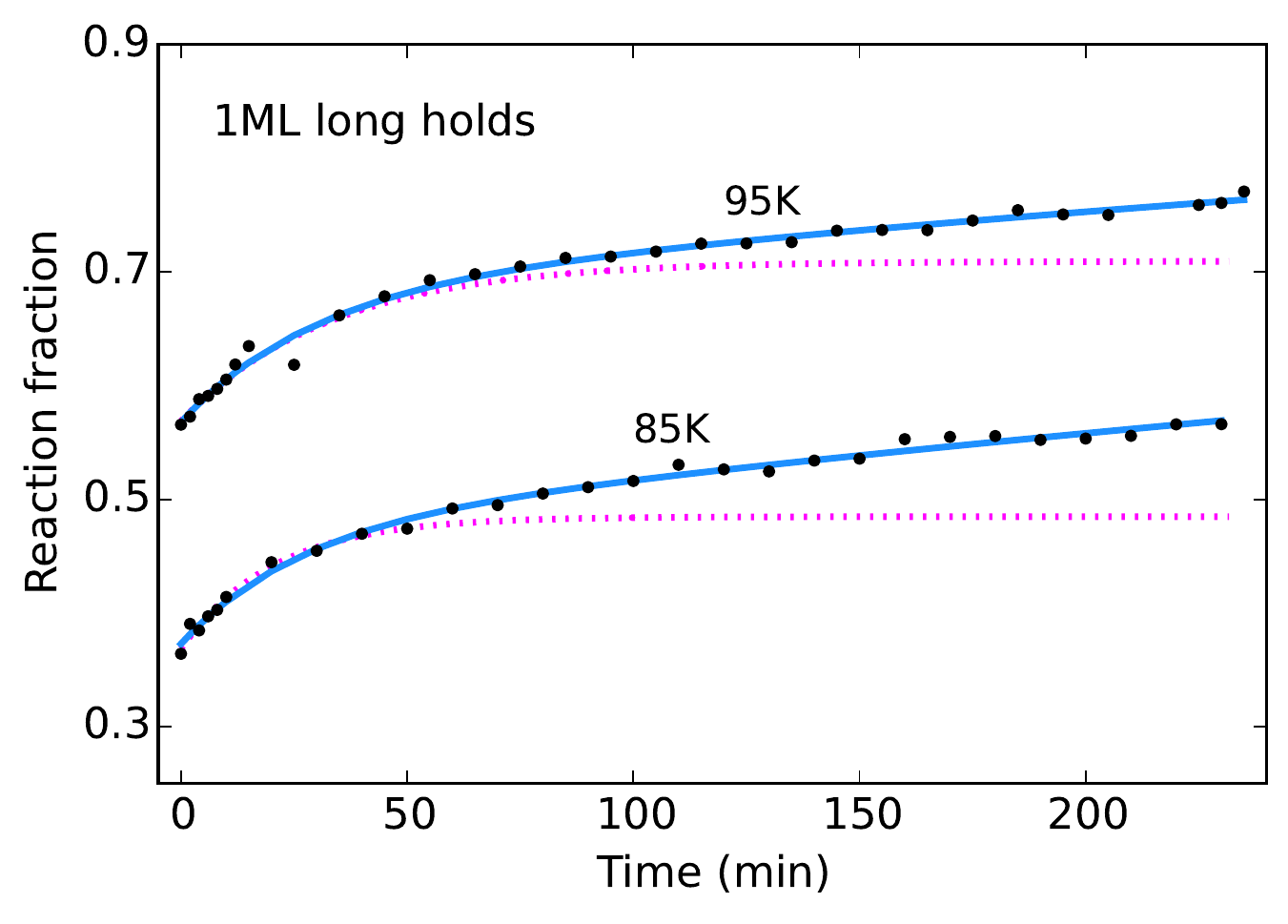}
		\caption{Isothermal salt growth from single monolayer experiments held at 85K (exp 11) and 95K (exp 12) are shown as black circles.  As for Figure \ref{1mlstep}, the first hour of reaction is fit with the single-step model (pink dotted lines). Blue lines show the two-step model fitting.}
		\label{1mllong}
	\end{figure} 
}
\def\placefiguremodelfits{
	\begin{figure}[h!]
		\centering
		\includegraphics[width=0.5\textwidth]{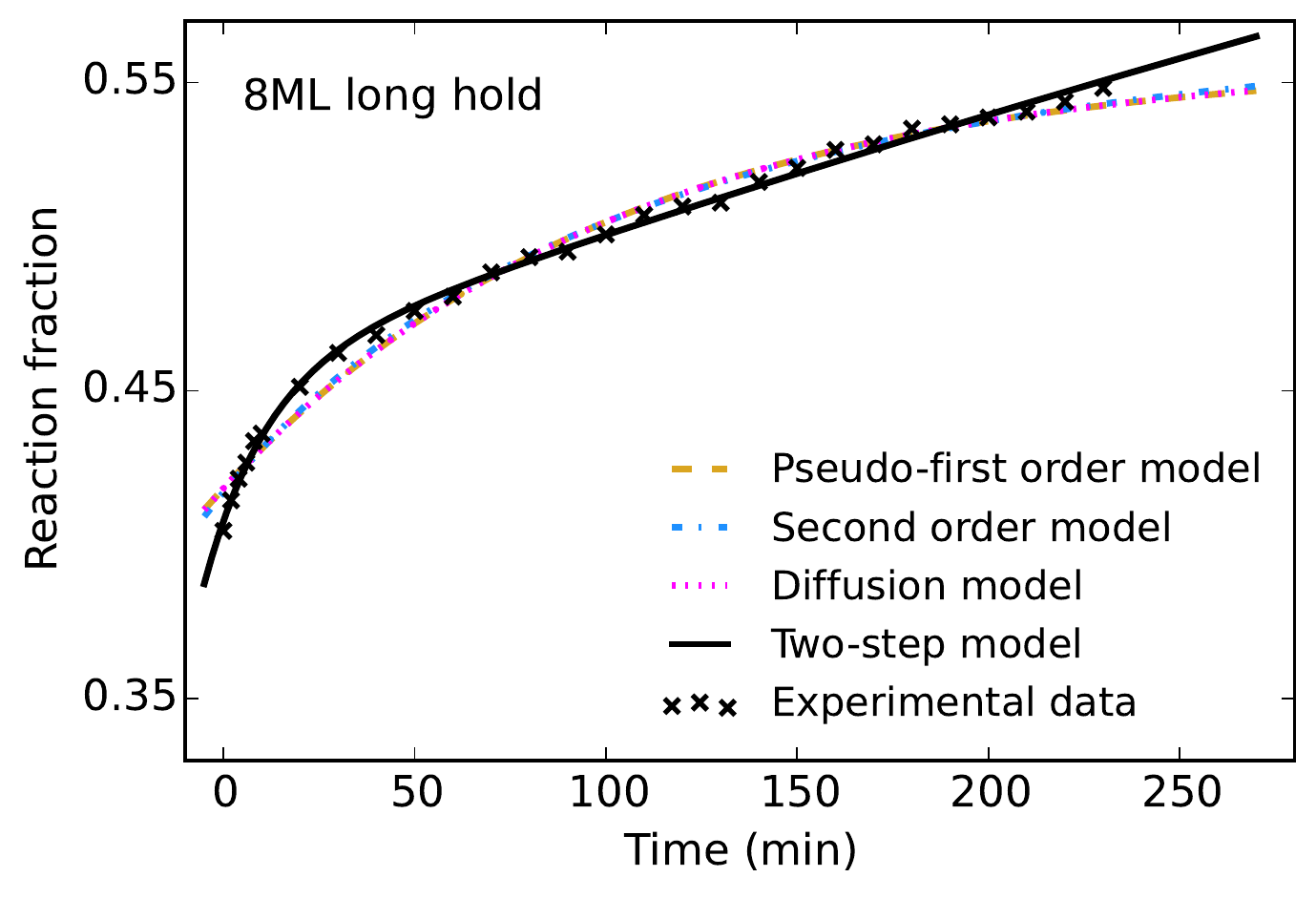}
		\caption{Growth curve for a thick long-hold experiment at 115K (experiment 19), fit by four different reaction models.}
		\label{modelfits}
	\end{figure} 
}
\def\placefigureArrhenius{
	\begin{figure}[h!]
		\centering
		\includegraphics[width=0.5\textwidth]{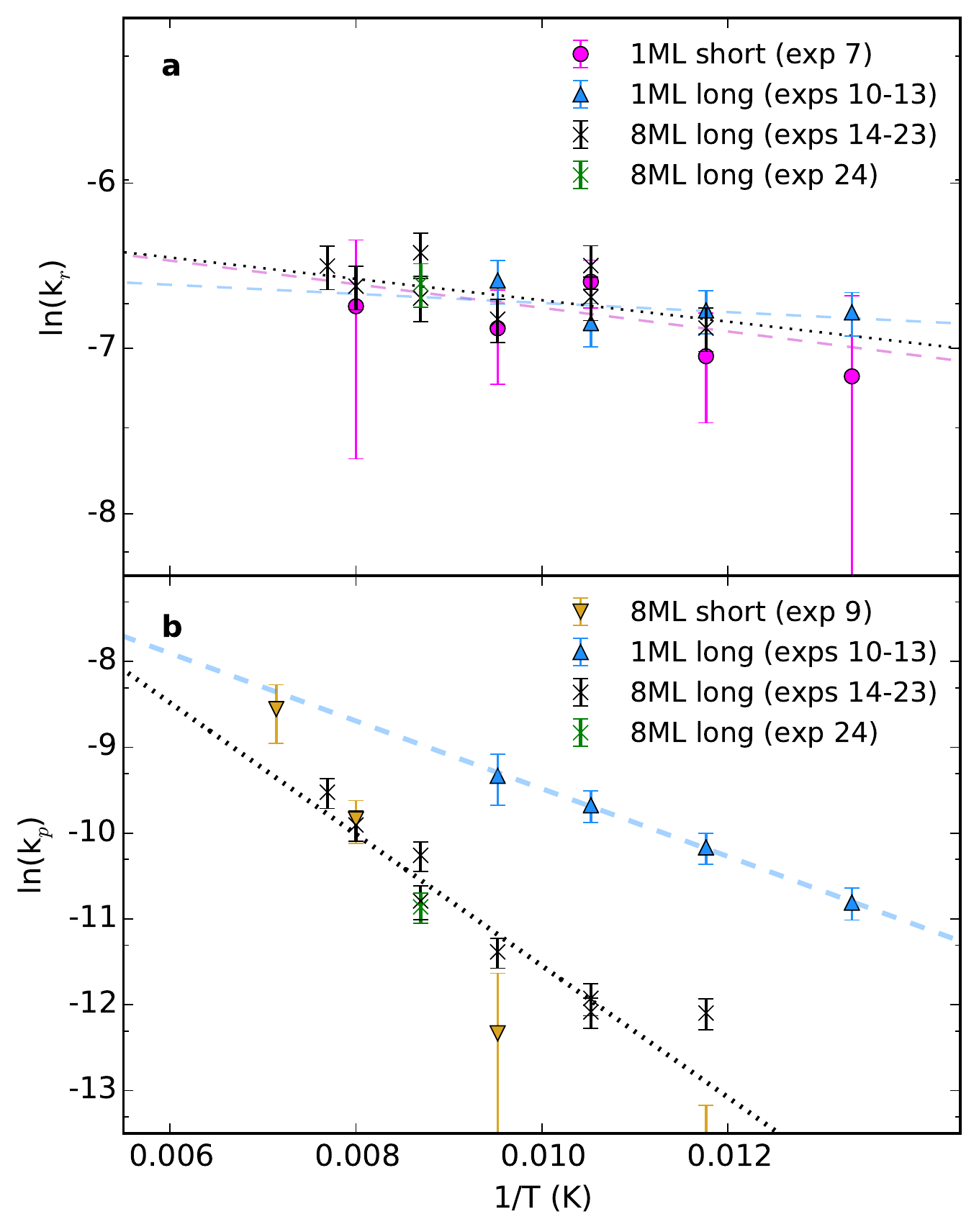}
		\caption{Arrhenius plot for the reaction barrier (a) and pre-reaction barrier (b), along with fits to each experimental category. Barriers for the 1ML short experiments (pink circles, panel a) were derived using a single-step reaction model; all other barriers are derived using a two-step model.  Of the 8ML long experiments, experiments with 4h isothermal holds (14-23) are shown with black x's, while the experiment with an 8h hold (24) is shown with green x's.}
		\label{Arrhenius}
	\end{figure} 
}

\def\placefiguresteps{
	\begin{figure}[h!]
		\centering
		\includegraphics[width=0.5\textwidth]{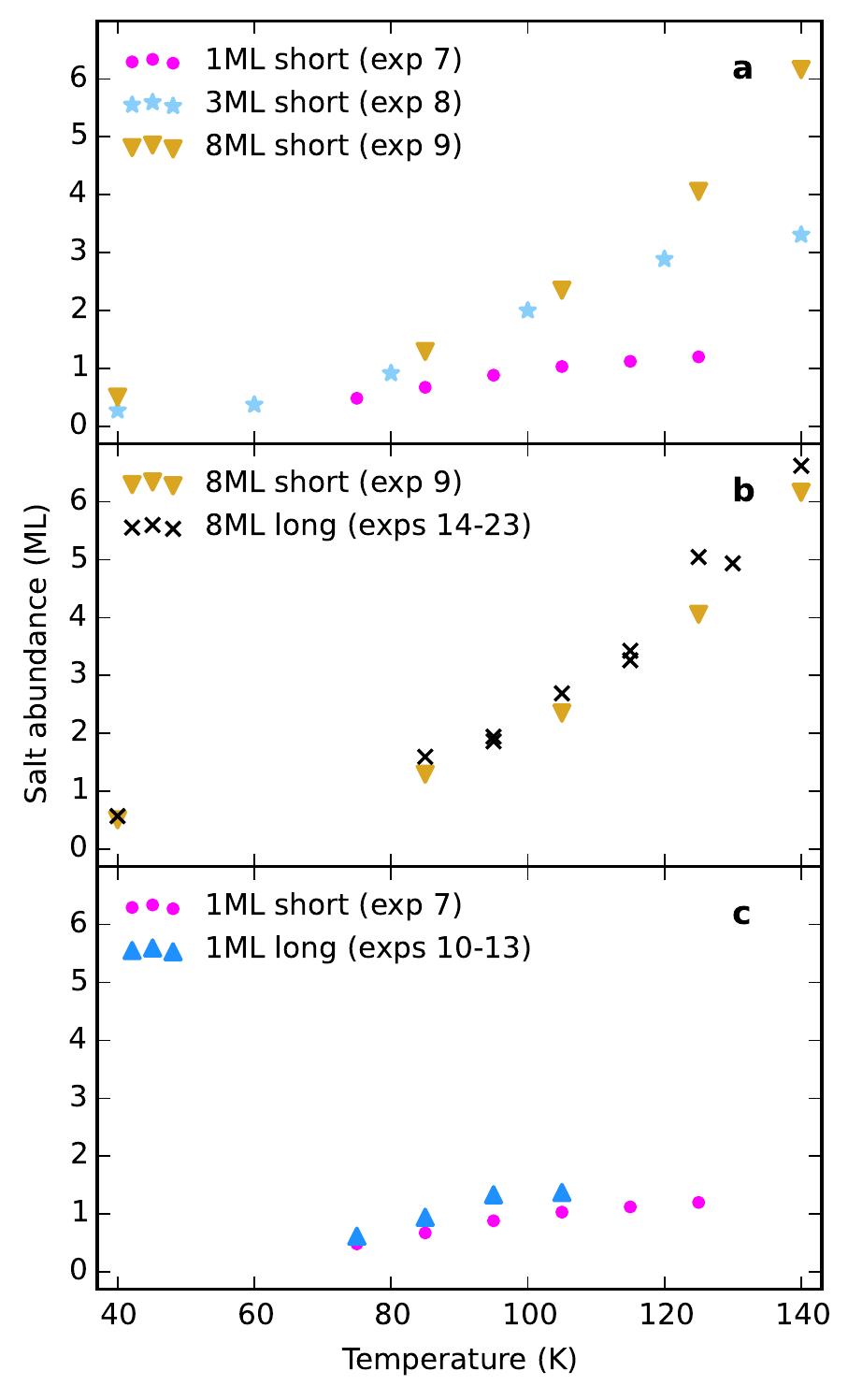}
		\caption{Salt growth (monolayers) that has occurred by the end of each isothermal growth period.  a: Different thicknesses of the step experiments (1h isothermal hold).  b-c: Step experiments (short holds) and separate experiments (long holds) for thick and thin ices.}
		\label{steps}
	\end{figure} 
}

\begin{abstract}
Interstellar complex organic molecules (COMs) are commonly observed during star formation, and are proposed to form through radical chemistry in icy grain mantles.  Reactions between ions and neutral molecules in ices may provide an alternative cold channel to complexity, as ion-neutral reactions are thought to have low or even no energy barriers.  Here we present a study of a the kinetics and mechanisms of a potential ion-generating acid-base reaction between NH$_{3}$ and HCOOH to form the salt NH$_{4}^{+}$HCOO$^{-}$.  We observe salt growth at temperatures as low as 15K, indicating that this reaction is feasible in cold environments.  The kinetics of salt growth are best fit by a two-step model involving a slow "pre-reaction" step followed by a fast reaction step.  The reaction energy barrier is determined to be 70 $\pm$ 30K with a pre-exponential factor 1.4 $\pm$ 0.4 x 10$^{-3}$ s$^{-1}$.  The pre-reaction rate varies under different experimental conditions and likely represents a combination of diffusion and orientation of reactant molecules.  For a diffusion-limited case, the pre-reaction barrier is 770 $\pm$ 110K with a pre-exponential factor of $\sim$7.6 x 10$^{-3}$ s$^{-1}$.  Acid-base chemistry of common ice constituents is thus a potential cold pathway to generating ions in interstellar ices.
\end{abstract}


\section{Introduction}
Complex organic molecules (COMs) have been detected towards a wide range of interstellar environments \citep{Herbst2009} and are thought to be the precursors to prebiotic molecules \citep[e.g.][]{Jørgensen2012,Belloche2013}. It is of particular interest to understand how they are formed and inherited through different stages of evolution in star-forming regions that may ultimately develop into solar systems capable of sustaining life. Current models of COM formation involve electron- or photon-induced dissociation of molecules in the icy mantles coating interstellar grain surfaces, producing radical species which diffuse and recombine to form larger molecules. This formation pathway requires temperatures exceeding $\sim$30K \citep{Garrod2006,Herbst2009}.  COMs were first detected in the cores of high-mass protostars \citep[e.g.][]{Blake1987,Helmich1997}, which undergo heating as they collapse and therefore experience temperatures above 30K; thus, these early detections are readily explained by the radical diffusion mechanism.  

Recently, COM detections in cold prestellar environments \citep[e.g.][]{Oberg2010,Bacmann2012,Cernicharo2012,Vastel2014} have challenged the established need for lukewarm ice chemistry, as there must be efficient cold channels to chemical complexity in order to produce COMs in these regions.  Several mechanisms, both gas-phase and grain-surface, have been suggested to explain COM production at low temperatures.  Based on observations of the B1-b core and the pre-stellar core L1544, respectively, \citet{Cernicharo2012} and \citet{Vastel2014} propose grain-surface formation of methanol and other smaller species, followed by nonthermal desorption and gas-phase reaction to form more complex molecules.  \citet{Balucani2015} have developed a gas-grain model to account for such a mechanism.  This model relies on efficient non-thermal desorption of methanol, followed by gas-phase radical-neutral and radical-radical reactions, and fairly well reproduces the observations of dimethyl ether and methyl formate towards L1544.  Ice chemistry alone could also be a viable pathway, provided that non-thermal diffusion is efficient, since it is the diffusion step that limits surface reactivity in low-temperature regimes.  For example, experiments involving UV photoprocessing of CH$_3$OH-rich ices have demonstrated efficient production of COMs at temperatures as low as 20K \citep{Oberg2009a}, and \citet{Oberg2010} suggest this mechanism to explain COM detections in the B1-b core.  \citet{Bacmann2012} similarly rely on non-thermal ice processing to explain the observed abundances towards the pre-stellar core L1698B: since it is well-shielded from UV radiation, they suggest that chemistry is likely induced by cosmic ray bombardment, secondary UV radiation from cosmic ray interactions with H$_{2}$, or energy from exothermic chemical reactions.  

Radical-radical reactions are not the only reactions without barriers, and reactions between ions and neutral molecules offer an alternative cold route to chemical complexity.  However, the importance of this channel is unknown since most previous work on COM formation has focused on radical reactions.  In the gas phase, ion-neutral reactions are responsible for many of the observed molecules in cold interstellar regions, and there may be an analogous grain-surface pathway.  Notably, recent theoretical work by \citet{Woon2011} demonstrates several such surface reactions to be barrierless.  Furthermore, compared to radical chemistry, ion chemistry does not rely on access to dissociative radiation.  While diffusion is still an obstacle for both ion chemistry and radical chemistry, it is important to understand their relative contributions to complex molecule formation.

Ions have been observed in ice mantles in a range of different interstellar environments, particularly OCN$^{-}$, and potentially also HCOO$^{-}$, and NH$_{4}^{+}$ \citep[see e.g.][]{Grim1987,Grim1989,Schutte1997,Schutte1999,Keane2001,Knez2005,Bisschop2007a,Oberg2011}.  Acid-base reactions are one potential source of ion generation in ice mantles.  This study will focus on the grain-surface reaction between NH$_{3}$ and HCOOH, which are among the most common constituents of ices in star-forming regions; observations indicate abundances with respect to water of 1-5\% for HCOOH \citep{Bisschop2007a, Boogert2008, Boogert2014} and 3-8\% for NH$_{3}$ \citep{Boogert2008, Bottinelli2010,Oberg2011, Boogert2014}. 

Similar acid-base systems have been studied experimentally in the past.  There have been several qualitative studies on the reaction of HNCO with NH$_{3}$, motivated by the identification of solid OCN$^{-}$ towards many different astrophysical objects. \citet{Raunier2003a} observe the proton transfer at 10K when HNCO is codeposited with an excess of NH$_{3}$ (1:10); however, HNCO deposited on top of NH$_{3}$ does not react until warmed to 90K.  To explain this, the authors perform quantum calculations and determine that the proton transfer is only spontaneous when the HNCO-NH$_3$ pair is stabilized by three or more solvating NH$_3$ molecules.    \citet{VanBroekhuizen2004} deposit mixed H$_2$O/NH$_3$/HNCO gases with varying H$_2$O concentrations and confirm that thermal processing is a robust mechanism for OCN$^-$ production.  They suggest that the reaction they observe at 10-15K may be due to kinetic energy brought by molecules when they freeze onto the surface, in addition to the solvation-induced reaction described in \citet{Raunier2003a}.  Subsequent reaction during warm-up is attributed to increasing NH$_3$ mobility.

In a quantitative study on the reaction between HCN and NH$_3$, \citet{Noble2013} co-deposit both species with an excess of NH$_{3}$.  The reaction is found to be thermally active, with some reaction occuring during deposition at 10K and further growth during warm-up.  They model isothermal growth curves as pseudo-first order with respect to HCN concentration and determine an activation energy of 324K. In another quantitative proton transfer study, \citet{Mispelaer2012} characterize and model the kinetics of the reaction between NH$_3$ and HNCO.  Fitting isothermal growth curves with rate equations, they find that it follows a two-step process: an initial, slow orientation step, followed by a fast reaction step, with an activation energy of 48K for the reaction.  They also perform fitting using a gamma-distribution of reaction constants to account for the fact that there will be a distribution of energy barriers to the orientation process, depending on the original position of different molecules.  Using this second method they determine an activation energy of 73K.  

There has been no quantitative study of the NH$_{3}$-HCOOH reaction at cryogenic temperatures; however, there have been several studies demonstrating that this reaction can occur at temperatures as low as 10K.  Early work by \citet{Schutte1997} was focused on spectroscopic assignments for the NH$_{4}^{+}$HCOO$^{-}$ ion pair, which was extended in \citet{Schutte1999} to demonstrate that in situ proton transfer occurs after deposition of an H$_2$O/NH$_3$/HCOOH mixture and the conversion increases during sample warm-up.  The kinetics are not quantified, but the authors note that the reaction seems to have a very small reaction barrier and that growth is limited by diffusion of reactant molecules. Later theoretical work by \citet{Park2006} suggests that the proton transfer is barrierless as long as at least three water molecules are present per reacting pair to stabilize the system.  More recently, experimental work by \citet{Galvez2010} confirms a thermally-induced in situ reaction of co-deposited NH$_{3}$ and HCOOH, with a small fraction (15\%) of reaction occurring upon deposition at 14K, and continuing as temperature is increased. 

The objective of this paper is to elucidate the kinetics and mechanisms of this process, and thus its feasibility and importance for the evolution of interstellar ices.  Experimental details are described in Section \ref{Experimental}, followed by the data analysis and modeling procedures in Section \ref{Analysis}.  The experimental results are presented in Section \ref{Results}, including the kinetic parameters extracted from the experiments as well as mechanistic inferences.  In Section \ref{Discussion} we compare our results with previous cryogenic acid/base studies and discuss the implications for surface channels to chemical complexity.

\section{Experimental Details}
\label{Experimental}
\subsection{Experimental Setup}
The experimental setup used for this experiment has been described previously by \citet{Lauck2015}.  Briefly, it consists of a CsI substrate window capable of being cooled to 11K by a closed-cycle He cryostat, with temperature monitored by a temperature controller (LakeShore 335) with an estimated accuracy of 2K and a relative uncertainty of 0.1K. The substrate is suspended inside an ultra-high vacuum chamber with a base pressure of $\sim$5x10$^{-10}$ Torr.  Ices are grown by introduction of gas vapors at a normal incidence through 4.8mm diameter pipes 0.7 inches from the substrate, unless otherwise noted.  

A Fourier transform infrared spectrometer (Bruker Vertex 70v) in transmission mode was used to measure the amount of each infrared-active species in the ice.  Gas-phase species present in the chamber were continuously monitored by a quadrupole mass spectrometer (Pfeiffer QMG 220M1).  The experiments were performed using NH$_{3}$ gas ($\geq$99.99\% purity, Sigma), HCOOH (98\%, Sigma), and deionized water (Sigma).  The HCOOH and water were purified using three freeze-thaw cycles with liquid nitrogen.
\subsection{Experimental Procedures}
\label{methods}
Table \ref{tab1} summarizes all experiments presented in this paper.  HCOOH and NH$_3$ were either co-deposited from separate pipes to form a mixed ice or sequentially deposited to form a layered structure.  All dosing took place at 14K.  Apart from the TPD experiments, the ices were heated at 5K min$^{-1}$ to a target temperature and maintained there for 1-4 hours while monitoring the ice composition, and finally the ices were heated to desorption.  IR scans taken at intervals of 10 minutes or less throughout the duration of each experiment.

\placetableone

Temperature programmed desorption (TPD) experiments (1-3) were performed for pure NH$_{3}$, pure HCOOH, and a layered NH$_3$/HCOOH ice.  By monitoring the masses of desorbing molecules and the temperatures at which they desorb, TPDs show chemical conversions that have occurred over the course of warm-up.  These experiments were performed by depositing $\sim$10 monolayers (ML) of each species and ramping the temperature at 5K/min until desorption (see section \ref{IRcalc} for how thickness is determined).  
\placefigureexperimental

For the co-deposition experiments (4-6), NH$_{3}$ and HCOOH were simultaneously deposited from two separate dosers at a distance of $\sim$1.2 inches and at an angle normal to the substrate.  The temperature was ramped at 5K/min to a specified temperature, held for 4 hours, and then ramped again at 5K/min to 240K.  The co-deposition at normal incidence resulted in a high degree of ice mixing, as evidenced by the large amount of salt growth.  A quantitative analysis of these experiments would require constraining the ice mixture homogeneity, which is beyond the scope of this study.

Most experiments (7-23) were performed using a layered ice configuration to reduce the amount of reaction that occurs during dosing.  Figure \ref{experimental} (top) shows a comparison of the co-deposited and layered experimental setups.  For all layered experiments, NH$_{3}$ was deposited first, followed by HCOOH.  The sample was heated at 5K min$^{-1}$ to a target temperature and held for a given amount of time, and then heated to 240K.

We perform two types of layered experiments: "step" and "single-temperature", shown schematically in Figure \ref{experimental} (bottom).  Step experiments are each held for 1h at several increasing temperatures within an experiment; thus, each step experiment yields several short isothermal growth curves at different temperatures.  Single-temperature experiments are each held for 4h at a single temperature per experiment, yielding a single long isothermal growth curve.  Experiments 7-9 are step experiments of different thicknesses: 1ML, 3ML, and 8ML.  Experiments 10-13 are single-temperature experiments of thin ices (1ML NH$_{3}$: 1ML HCOOH), and experiments 14-23 are single-temperature experiments of thick ices (9ML NH$_{3}$: 7ML HCOOH).  Experiment 24 is a single-temperature experiment of a thick ice, with an 8 hour isothermal hold instead of 4 hours in order to verify that the growth curve is sufficiently sampled within a 4 hour timescale.  By varying the conditions of thickness and time, we can assess the salt growth kinetics under different experimental conditions and thereby better constrain the mechanism of reaction.  

\section{Analysis and Modeling}
\label{Analysis}
\subsection{IR spectra}
\label{IRcalc}

Concentrations of ice species of interest were determined from baseline-subtracted infrared spectra.  Each spectrum is averaged over 128 interferograms and takes approximately 2 minutes to complete.  Ice thickness was calculated from the formula:
\begin{equation}
N_{i} = \frac{2.3\int\tau_{i}(\nu)d\nu}{A_{i}} \label{col_density}
\end{equation}
where $N_{i}$ is column density (molecule cm$^{-2}$), $\int\tau_{i}(\nu)d\nu$ is the integrated area of the IR band (absorbance units), and $A_{i}$ is the band strength in optical depth units.  The standard monolayer coverage of 10$^{15}$ molecules cm$^{-2}$ was assumed.

The band positions and strengths used to determine the thickness of NH$_{3}$ and HCOOH are taken from \citet{Bouilloud2015} and were chosen to be the strongest features that do not overlap with the growing salt features (Table \ref{Tab2}).  Band strength uncertainties of 20\% will result in the same uncertainty for the measured ice thicknesses listed in Table \ref{tab1}.

\placetabletwo

IR scans are taken throughout the course of an experiment, enabling in situ monitoring of the concentrations of ice species.  These ice concentrations are plotted as a function of time over the course of an experiment to yield growth curves.

\subsection{Growth curve fitting}
In order to extract kinetic parameters, isothermal growth curves are fit according to rate equations.  To measure the reaction rate ideally, growth would depend only on the barrier to proton transfer; in reality, however, it is likely that an ensemble of diffusion and reorientation processes ("pre-reaction" steps) will inhibit growth.  Because it is not clear which of these will contribute under different experimental conditions, we have used several different models to see which provides the best fit to the experimental growth curves.  We first give a brief overview of the kinetic modeling, and then describe the four models used in this work.

The general formalism for reaction kinetics is: 
\begin{equation}
\frac{dX}{dt} = k(T)f(X) \label{gen_kin}
\end{equation}
where $X$ is the reaction fraction, $k$ is the rate constant,  $T$ is temperature (K), and $f(X)$ is the reaction model.  The reaction fraction is determined by normalizing each growth curve to the final amount of salt at the completion of the reaction (ie, just prior to desorption).  The rate constant can in turn be expanded to
\begin{equation}
k(T) = Ae^{-E_{a}/T} \label{arrhenius}
\end{equation}
where $A$ is the Arrhenius prefactor and $E_{a}$ is the activation energy.  For solid-state reactions, the choice of reaction model $f(X)$ is not obvious since local inhomogeneities and diffusion effects preclude the kinetics from being described by concentration-dependent rate-laws used for homogeneous fluid systems.  However, order-based methods are still useful as empirical models that allow kinetic parameters to be extracted from experiments \citep{Vyazovkin1997,Khawam2006}.  Recently, order methods have been successfully applied to a number of thermal reactions of astrochemical relevance \citep{Bossa2009a,Bossa2009,Theule2011,Mispelaer2012,Noble2013}. For the reaction studied here:
\begin{equation}
\mathrm{NH_{3} + HCOOH \rightarrow NH_{4}^{+} + HCOO^{-}} \label{rxn}
\end{equation}
we define the change in reaction fraction 
\small
\begin{equation}
\frac{dX}{dt} = \frac{d(\mathrm{HCOO}^{-})}{dt} = \frac{d(\mathrm{NH}_{4}^{+})}{dt} = \frac{-d(\mathrm{HCOOH})}{dt} \label{fractions}
\end{equation}
\normalsize
where parentheses denote the fraction of each individual species relative to its maximum value in the course of the experiment.  The fractional changes in concentration of HCOO$^-$, NH$_4^+$, and HCOOH can be equated since each have a maximum value equal to the initial value of HCOOH, the limiting reactant.  The reaction order model is then: 
\begin{equation}
\frac{dX}{dt} = k(T)(\mathrm{NH}_{3})^{\alpha}(\mathrm{HCOOH})^{\beta} \label{gen_order}
\end{equation} In this work, growth curves were fit with different variations of this general model in order to evaluate different mechanisms by which the reaction may take place.

\textit{Pseudo-first order:}  Because NH$_{3}$ in in excess of HCOOH for the majority of experiments, the simplest kinetic model would be pseudo-first order with respect to HCOOH: 
\begin{equation}
\frac{dX}{dt} = k(T)(\mathrm{HCOOH}) \label{pseudo_first}
\end{equation}

\textit{Second order:}  The next step in complexity is if the reaction is first-order with respect to both HCOOH and NH$_{3}$, for a total order of two: 
\begin{equation}
\frac{dX}{dt} = k(T)(\mathrm{NH_{3}})(\mathrm{HCOOH}). \label{first_first}
\end{equation}  
Modeling the reaction as second-order with respect to either reactant did not improve the fit, and no higher-order processes were considered.

Both \eqref{pseudo_first} and \eqref{first_first} are single-step processes that consider only the reaction rate.  We also tested a single-step model that assumes diffusion-regulated kinetics, and a more complex model that accounts for both mobility and reaction kinetics potentially at play in the solid-state environment. 

\textit{Diffusion model:}  A Fickian diffusion model was adapted (see \citet{Lauck2015} for details) to describe the layered system of NH$_{3}$ beneath HCOOH.  In this model, the NH$_{3}$ is assumed to be the mobile diffusant into a matrix of immobile HCOOH.  The mixed fraction is described by: 
\footnotesize
\begin{equation}
N_{mix}(t) = \frac{N_o(h-d)}{h} - \sum_{n = 1}^{\infty}\frac{2N_oh}{n^{2}\pi^{2}d}\sin^{2}(\frac{n\pi d}{h}) \exp[{\frac{-n^{2}\pi^{2}}{h^{2}}D(t + t_0)}]
\end{equation}
\normalsize
\noindent
where $h$ and $d$ are, respectively, the total ice height and the height to the interface (both in nm), $D$ is the diffusion constant (cm$^{2}$ s$^{-1}$), $N_o$ accounts for thickness uncertainties, and $t_0$ is a time offset.  This model assumes that the kinetics are determined entirely by diffusion, and that reaction occurs immediately; in other words, the rate of mixing entirely determines the rate of salt growth.

\textit{Two-step model:}  The final model used was a two-step model involving a slow pre-reaction step followed by a fast reaction step.  A similar treatment was used by \citet{Mispelaer2012} in fitting the reaction of NH$_{3}$ with HNCO.  Several mechanistic possibilities exist for the pre-reaction step; as we discuss in more detail in the following section, it likely consists of both orientation and diffusion, with different processes dominating the pre-reaction kinetics under different conditions.  Because of this mixed nature, the pre-reaction step may depend on the temperature, ice structure, and elapsed time of a given experiment.  The two-step model allows us to absorb the pre-reaction processes into the first step and therefore isolate the actual reaction step.

Here, we express the slow pre-reaction step as a unimolecular conversion of HCOOH from an inactive form to an active form.  We write this step as:
\begin{equation}
\mathrm{HCOOH}^{o} \rightarrow \mathrm{HCOOH}^{*}
\end{equation}
with a rate $k_{p}$, where $^{o}$ and $^{*}$ represent the inactive and active form, respectively. The fast reaction step that follows is then: 
\begin{equation}
\mathrm{NH}_{3} + \mathrm{HCOOH}^{*} \rightarrow \mathrm{NH}_{4}^{+} \mathrm{HCOO}^{-}
\end{equation}
with a rate $k_{r}$.  We can express the kinetics of the total process in terms of the fractional conversion of HCOOH: 
\begin{eqnarray}
\frac{d(\mathrm{HCOOH}^{o})}{dt} = - k_{p}(\mathrm{HCOOH}^{o}) \\
\frac{d(\mathrm{HCOOH}^{*})}{dt} = k_{p}(\mathrm{HCOOH}^{o}) - k_{r}(\mathrm{HCOOH}^{*})
\end{eqnarray}
This system of equations is solved for the concentrations of both HCOOH$^{o}$ and HCOOH$^{*}$ as a function of time; since the two are indistinguishable by IR spectroscopy, our observable quantity (the total rate of consumption of HCOOH) is the sum of these two contributions.  Recalling equation \eqref{fractions}, we derive an expression for the reaction fraction $\alpha$ as a function of time:
\begin{equation}
X(t) = \frac{e^{-t(k_{p} + k_{r})}[e^{k_{p}t}(k_{p} - k_{r} c)  - e^{k_{r}t}k_{r}(1 - c)]}{k_{p} - k_{r}}
\end{equation}
where $c$ is the initial fraction of HCOOH in the active form, or (HCOOH$^{*})_{t=0}$.

Because kinetic fits are performed with respect to the reaction fraction, any uncertainties in the absolute band strength of the salt will not impact the kinetic parameters derived from the growth curves, as we are fitting a ratio.  However, because the final amount of salt is measured at a different temperature than the growth curves, uncertainties in the temperature dependence of the band strength may introduce error into the fitting (see Section \ref{uncert} and Appendix I).

All modeling is done only for isothermal growth curves.  We do not attempt to model salt growth that occurs during deposition, as this may be governed by a different mechanism.  For instance, it is possible that a "hot molecule" mechanism by which energy dissipated by gas-phase molecules is in part responsible for the initial salt formation that occurs during deposition.

\subsection{Uncertainty analysis}
\label{uncert}
We consider five sources of error: spectral line fit errors, band strength uncertainties, kinetic fit uncertainties, dispersion between identical experiments, and absolute temperature calibration errors.  The spectral line fit uncertainty consists of both gaussian fit uncertainty and uncertainty due to baseline selection.  The former is very small, with errors generally at least 4 orders of magnitude smaller than the values of the peak areas.  The latter is thickness-dependent and results in uncertainties of 0.5\%-1\% in the peak areas of measured salt features.  This source of uncertainty dominates the error bars for individual measurements of IR peak area.

Uncertainties in the IR band strength of the measured salt feature will not contribute to the uncertainty for an individual fit of a growth curve, as we fit the reaction fraction rather than the total number of monolayers of salt.  Thus, any uncertainty in the absolute value of the band strength will cancel out.  However, the band strength likely varies with temperature and this must be considered when comparing salt growth in experiments run at different temperatures.  We assume a 5\% uncertainty in the magnitude of growth curves due to this temperature dependence (see Appendix I for a detailed explanation of band strength uncertainty).  This is the most important contribution to the error bars on rate constants for individual experiments.

Statistical errors from fitting the kinetic growth curves are very small and do not contribute significantly to the rate constants uncertainties (less than 1\%).

Dispersion measurements are determined using identical experiments to incorporate variations in chamber and ice morphology that cannot be directly controlled for. This is the main source of uncertainty when comparing multiple experiments, with a 15\% uncertainty for reaction rate constants and 17\% uncertainty for pre-reaction rate constants which are propagated into fitting for the reaction barriers and pre-exponential factors.

The error from calibration of the absolute temperature reading will contribute only to the derived energy barriers and pre-exponential factors.  However, this uncertainty is very small compared to the dispersion error.

\section{Results}

\label{Results}
The experimental results are presented in the following order: IR spectra in \ref{sec_IRspec}, TPDs in \ref{sec_TPD}, co-deposited experiments in \ref{sec_codep}, qualitative analysis of layered experiments in \ref{qual}, growth curve fitting in \ref{sec_modeling}, reaction barrier determination in \ref{sec_kr}, and pre-reaction barrier determination in \ref{sec_kp}.  It should be noted that the ultimate aim is to determine the reaction barrier of the proton transfer, which is necessarily tangled with other processes due to the limits of experimental constructs.  We attempt to constrain pre-reaction steps as well as possible in order to isolate the reaction barrier, but the kinetic parameters derived for the pre-reaction are not necessarily relevant to astrophysical conditions since our system consists of simple, well-controlled ice structures rather than a more complicated and water-dominated mixture.

\subsection{IR spectra}
\label{sec_IRspec}
Figure \ref{IR}a shows the IR spectra for pure NH$_{3}$, pure HCOOH, and a layered NH$_3$/HCOOH mixture at increasing temperatures.  Salt features, identified as those that grow over the course of the reaction, are consistent with those reported by \citet{Galvez2010} and are listed in Table \ref{Tab2}. Small salt features are apparent at 15K, with slow growth until 80K and rapid growth beyond this.  Desorption is complete by $\sim$220K.  
\placefigureIR

The four most prominent peaks associated with salt growth occur within the spectral window of 1300cm$^{-1}$ to 1600cm$^{-1}$; while there is some overlap of especially the inner two peaks, a Gaussian fitting of the four peaks produces an excellent fit to the overall shape of the spectrum, as shown in Figure \ref{IR}b.  The fitting procedure uses the spectral window from $\sim$1320-1600 cm$^{-1}$, excluding the tail from 1600-1750 cm$^{-1}$ which is due to the pure HCOOH feature at 1710 cm$^{-1}$ and the pure NH$_3$ feature at 1625 cm$^{-1}$.  Regions outside of this window are assigned a value of zero to allow the model to return to the baseline.  The growth curves derived from the areas of the two outermost peaks at 1346cm$^{-1}$ and 1573cm$^{-1}$ are in very good agreement with one another, providing confirmation that these peaks are indeed originating from the growth of the same species.  The inner two peaks follow less consistent growth curves, possibly due to their greater degree of overlap.  In addition, the NH$_{4}^{+}$ $\nu_{4}$ band is very broad and it is difficult to distinguish an exact peak position. The HCOO$^{-}$ $\nu_5$ band at 1377cm$^{-1}$ may also have a contaminating component from the pure HCOOH band at 1380cm$^{-1}$, although this is weaker by an order of magnitude \citep{Bouilloud2015}.  The $\nu_{2}$ formate band at 1573cm$^{-1}$ is used for all growth-curve fitting since it is the most consistent and prominent salt feature. Band strengths for the salt IR modes were derived for each of the four peaks and are listed in Table \ref{Tab2}.  Details of their derivation are given in Appendix 1, along with an analysis of structure and temperature dependencies.

\subsection{TPDs}
\label{sec_TPD}
Temperature programmed desorption experiments were performed for pure NH$_{3}$, pure HCOOH, and a layered experiment of NH$_3$ under HCOOH.  The ices were heated at 5K/min until desorption while continuously monitoring gas-phase concentrations with the QMS.  Figure \ref{TPDs} shows the resulting QMS traces for these experiments.  Mass 29 traces the dominant HCOOH fragment, and mass 17 traces NH$_3$.  The mixture of NH$_{3}$ and HCOOH desorbs at higher temperatures than either pure ice, indicating a chemical conversion has indeed occurred (peak desorption temperatures for NH$_3$, HCOOH, and the salt are 96K, 144K, and 203K, respectively).  Furthermore, some NH$_{3}$ desorbs prior to salt desorption, but HCOOH desorbs only as the salt; this indicates that the HCOOH has undergone complete conversion, and in experiments with excess NH$_{3}$, the total salt formation will be determined by the initial dose of HCOOH. 
\placefigureTPDs

We note that mass 17 also traces a fragment of H$_2$O in addition to the major NH$_3$ channel.  We expect that water will be a minor contribution to the total QMS signal since the samples are formed from high purity gases maintained under ultra-high vacuum.  However, there may be background deposition of water onto the sample throughout the experiment and thus some degree of this signal could be due to water.  Based on an analysis of the dominant water channel at mass 18, around the water desorption temperature of 160K, water contamination contributes up to 20\% of the total mass 17 signal; at all other temperatures, the water contamination contributes 10\% or less to the measured mass 17 signal.

\subsection{Salt formation in co-deposited ices}
\label{sec_codep}
\placefigurecodep
Growth curves for co-deposited experiments are shown in Figure \ref{codep}. For all co-deposited experiments, considerable salt growth occurs immediately following dosing at 14K ($\sim$30\% of the total salt conversion), with further growth during subsequent warm-up from 14K to the target temperature.  For the target temperature of 85K, growth during the isothermal hold is evident; however, no isothermal growth was seen for the co-deposited experiments with 20K and 65K target temperatures.  After the isothermal holds, the temperature was increased until desorption, and growth continued once again during this final warm-up.  The rapid growth during the initial warm-up followed by a lack of growth at target temperatures below 85K suggests that the kinetics of salt formation under these conditions are complex: growth is evidently accessible at low temperatures as long as the sample is being warmed, but during isothermal periods the growth is arrested unless the temperature is sufficiently high.

\subsection{Qualitative analysis of salt growth in layered ices}
\label{qual}  
A comparison of layered and co-deposited experiments, both with 85K target temperatures, is shown in Figure \ref{colay}.  Following deposition, both experiments exhibit similar patterns of salt growth over the course of the experiment.  However, a layered setup offers the advantage of greatly reducing the amount of salt formation that occurs during dosing ($\sim$0.5ML, compared to $\sim$1.8ML in co-deposited ices).  This enables a better estimation of the amount of available reactants compared to the co-deposited case.  Additionally, it minimizes the amount of salt that has formed during the warm-up to the targeted temperature, enabling a better evaluation of the early kinetics of the reaction.  Because of this, we performed our analysis using a layered ice setup, rather than co-deposition as is used by many previous kinetic studies \citep[e.g.][]{Mispelaer2012,Noble2013}.
\placefigurecolay

We note that, while the layered experiments exhibit a lower absolute growth, it is in fact a higher reactivity per molecule in contact: $\sim$50\% salt conversion occurs in the 1ML HCOOH in contact for the layered experiment, compared $\sim$25\% salt conversion for the 7ML HCOOH in contact in the co-deposited experiment.  It is possible that the salt formation during deposition may be occurring by different mechanisms in the co-deposited and layered experiments.  For instance, a hot-molecule mechanism might contribute to growth preferentially in the layered experiment.

\placefiguresteps

We first look qualitatively at the amount of salt that has formed at the end of each isothermal hold for the different categories of experiments described in Section \ref{methods}.  Figure \ref{steps}a shows the growth differences between step experiments of ices with different thicknesses.  Above 80K, the 3ML and 8ML experiments produce more salt than the 1ML experiment.  In this regime, diffusion must occur since non-interface molecules contribute to salt growth in the multilayer ices.  Furthermore, the 3ML and 8ML experiments grow identically until about 120K, at which point the 8ML ice growth continues faster than the 3ML growth.  This can be attributed once again to diffusion: while short-range diffusion appears to contribute above 80K, long-range diffusion is important only above 120K.  

In Figure \ref{steps}b-c, the 8ML and 1ML step experiments are compared to the single-temperature experiments with the same thicknesses.  Regardless of the fact that prior reaction has occurred in the step experiments, roughly the same amount of salt is produced by a given temperature as for the single-temperature long-hold experiments.  This has several implications, notably that most salt growth occurs in the first hour at a given temperature, and that at each temperature the salt formation kinetics does not depend on how pre-existing salt has formed.  We discuss this observation in more detail in Section \ref{Discussion}.

\placetablefour

\subsection{Growth curves and model fitting}
\label{sec_modeling}
In fitting the kinetics of salt formation, we do not attempt to model growth that occurs during deposition or temperature ramps.  Our modeling of isothermal growth takes into account previous salt formation, however, and so this warm-up growth should not be problematic to our kinetic derivations.  Previous works \citep{Mispelaer2012,Noble2013} offset the reaction fraction to equal zero at time zero, which models salt growth as if there is no salt present in the ice at the start of the isothermal period.  Instead, we fit for a time offset that encompasses any salt growth prior to time zero of the isothermal hold.  This allows us to account for prior salt growth regardless of its origin, i.e. whether it formed during deposition, warm-up, or isothermal growth.  All resulting reaction constants ($k_r$) and pre-reaction constants ($k_p$) are summarized in Table \ref{ks}.

\textit{Thin short experiments: }
The layered step experiment of $\sim$1ML NH$_3$ under 1ML HCOOH is shown in Figure \ref{1mlstep}. The salt grows rapidly during warm-up periods, and isothermal periods exhibit slower growth.  Isothermal holds at 85K, 95K, and 105K produce the most salt growth; the kinetics are likely quite slow at 75K, and above 105K the reactants are mostly consumed and of limited availability, thus slowing growth. The growth curves are well-fit by the pseudo-first order model, and more complex models do not improve the fit.  Indeed, the diffusion model does not converge on a solution at all, which is not surprising given that diffusion should not contribute much to growth in single-ML ice layers.  The two-step model produced negative values for k$_r$, indicating that in the case of short reaction timescales and thin ices, salt growth is determined by only the reaction barrier.  In other words, under these conditions the energy barrier of the proton transfer dominates the kinetics, and dynamical processes do not play an important role.
\placefiguremlstep

\textit{Thin long experiments: }The growth curves for single-temperature experiments of 1ML ices are shown in Figure \ref{1mllong}.  Early reaction times were fit with the single-step model as was done for the step experiments of 1ML ices.  We expect that if a single kinetic process contributes to growth, then the same model should work at early and late timescales.  However, it is clear that different kinetics contribute at early and late times, and a model of early growth cannot reproduce late growth.  In the limit of longer experiments, salt growth is instead best modeled by two-step kinetics, dependent on both the reaction barrier and a pre-reaction barrier.  This suggests that while the growth seen in the shorter experiments is dominated by reaction between molecules already in an active state, in the limit of longer time when these active molecules are fully consumed, the contribution of the slow pre-reaction step becomes evident.  Because each ice layer is $\sim$1ML thick, it is unlikely that diffusion is the important pre-reaction step in this case.  Instead, we propose a re-orientation of reactant molecules into a favorable reaction configuration, similar to that described in \citet{Mispelaer2012}.  
\placefiguremllong

\textit{Thick long experiments: }The 8ML step and single-temperature experiments were initially fit with all four kinetic models; an example growth curve is shown in Figure \ref{modelfits}.  Again the two-step model produces an excellent fit, while the other models do not capture the shape of the growth curve.  In this case, the pre-reaction step is likely a combination between diffusion and orientation since, as discussed previously, diffusion is seen to contribute to salt growth in multilayer ices.

The pre-reaction barrier becomes an important contribution to growth on long timescales; thus, to verify that it is being sufficiently sampled on 4 hour timescales, we performed an experiment with an 8 hour isothermal hold.  Comparing experiment 24 with the 4hr experiments 19 and 20 in Table 3, we find that both the reaction and pre-reaction barrier for the 8hr experiment are consistent with the values extracted from the 4hr experiments.  The 4 hour growth curves are therefore sufficient for use in fitting, as the inferred kinetics do not change on longer timescales.  The two-step model predicts that the growth curve should approach a stable value equal to a reaction fraction of 1 on the timescale of tens of hours.
\placefiguremodelfits

\subsection{Reaction barrier determination}
\label{sec_kr}
The reaction rate constants derived from the growth curve fitting were used to generate an Arrhenius plot (Figure \ref{Arrhenius}a).  Fitting all data points simultaneously, we derive a reaction energy barrier $E_{a}$ = 70 $\pm$ 30 K and pre-exponential factor $A$ = 1.4 $\pm$ 0.4 x 10$^{-3}$ s$^{-1}$.  Note that rate constants from below 70K and above 130K are excluded in this fitting, as the reaction is too slow to fit growth at low temperatures and most of the reactants have already been consumed by high temperatures. The error bars for thin ices are much larger than for the thick ices, as it is more difficult to constrain the fit when the magnitude of growth is smaller, and especially so when the timescale is short as for the step experiments.  

The reaction rate constants for the thin short, thin long, and thick long experiments at the same temperature are in very good agreement, and fall within each others error bars as seen in Figure \ref{Arrhenius}a.  Likewise, when each category is fit individually, the derived values for E$_a$ and A are all consistent within the combined uncertainties (Table \ref{Eas}).  Thus, despite differences in thickness and timescale, we still find consistent values for the reaction barrier using the two-step model, indicating that we are in fact isolating the reaction step from the pre-reaction processes.
\placefigureArrhenius

\placetablethree

\subsection{Pre-reaction barrier determination}
\label{sec_kp}
A similar treatment can be performed on the pre-reaction constants as is described above.  Unlike for the reaction barrier, each category of experiments is likely to have a different pre-reaction barrier depending on the experimental conditions; in other words, orientation and diffusion are expected to contribute to growth to different extents between setups. 

Fitting the thick long (single-temperature) experiments results in a pre-reaction activation barrier of 950K with a pre-factor of order 0.04; however, the uncertainties on these numbers are very large.  To better constrain the pre-reaction barrier, we have re-fit all experiments with a fixed reaction constant $k_{r}(T)$ based on the combined $E_{a}$ and $A$ values in Table \ref{Eas}.  This is especially important for experiments with thin ices or short timescales since these fits are more uncertain.  The resulting fits to growth curves using the fixed parameter method are in excellent agreement with the data.  The pre-reaction rate constants from the fixed fitting reaction kinetics are shown in Figure \ref{Arrhenius}b, and the resulting kinetic parameters are listed in Table \ref{Eas}.

The pre-reaction rates from thick experiments with short and long isothermal holds appear reasonably consistent, and Arrhenius fitting results in a much steeper slope than for the thin experiments, corresponding to a higher energy barrier.  We expect the pre-reaction process to be dominated by diffusion for the thick ices and by orientation for the thin ices, and it is not surprising that orientation would have a lower energy barrier than diffusion.  Additionally, the pre-reaction barrier for the thick ices appears to have a tail at the 85K and 95K temperature points.  Based on the qualitative analysis of Figure \ref{steps}a, diffusion appears to contribute to multilayer growth above 80K; it is likely that the shallower slope of the Arrhenius plot at lower temperatures corresponds to an orientation-dominated pre-reaction step, and the steeper slope at higher temperatures represents a diffusion-dominated pre-reaction step.

\section{Discussion}
\label{Discussion}
\subsection{NH$_3$ + HCOOH reaction}
\label{rxndisc}
We extract an energy barrier of 70K for the proton transfer between NH$_3$ and HCOOH using the isothermal rate constants from different experiments (Figure \ref{Arrhenius}).  This barrier is consistent for different experimental conditions, demonstrating that it is indeed isolated from the pre-reaction processes that depend on the experimental construct.

The salt growth observed in this work is consistent with the qualitative description by \citet{Schutte1999}, who observed some reaction upon deposition at 10K and further growth during warm-up.  The authors speculated that the reaction would have a low barrier and be limited by the diffusion of reactants, both of which we have demonstrated here.  Indeed, we find that not only diffusion but also an orientation of reactant molecules contribute to the overall growth rate of the salt. 

Comparing our results with previous quantitative proton transfer studies, the reaction barriers are quite similar, despite differences in fitting methodologies: \citet{Noble2013} derive the reaction barrier for NH$_3$ + HCN to be 324K, \citet{Mispelaer2012} find the barrier for HNCO + NH$_3$ to be 48K, and we find the barrier for NH$_3$ + HCOOH to be 70K.  The HCN barrier perhaps appears high in comparison, but HCN is also a far weaker acid than HCOOH in the aqueous phase. 

We note three differences with these existing studies that may impact the derived kinetic parameters.  First, \citet{Mispelaer2012} use a two-step model as in this work, but \citet{Noble2013} use a single-step pseudo-first order model; if the reaction indeed follows two-step kinetics, a single-step fit would likely result in too high of a reaction barrier since the slow pre-reaction step is also being incorporated into the fit.  This may contribute to the relatively high barrier found for the HCN + NH$_3$ reaction.  Second, both previous studies analyzed reaction in co-deposited ices, whereas our ice is layered.  This introduces diffusion as an important pre-reaction process in our salt growth, whereas orientation may be the most important pre-reaction step in previous works.  Finally, the previous works perform fitting on growth curves that are normalized such that at time zero the pure ice has a fraction of unity and the product has a fraction of zero.  This method assumes that any reaction that has already occurred during the warm-up phase of the experiment does not influence the kinetics during the isothermal period.  In this work, we do not offset our growth curves to begin at a reaction fraction of zero because prior reaction could inhibit further reaction and should therefore be taken into account in the isothermal modeling.  

\subsection{Pre-reaction processes}
Unlike the reaction barrier, the derived pre-reaction barriers differ between experimental setups, reflecting the different dynamical processes at play under different experimental conditions.  Re-orientation of reactant molecules into favorable reaction configurations should occur in all ices; this process contributes to growth once molecules that are already oriented to react are consumed.  Diffusion of reactant molecules occurs only in thick ices, with molecules not originally at the interface replenishing the reactive stock.  We find diffusion-limited (thick) ices to have much higher pre-reaction barriers than orientation-limited (thin) ices, which is consistent with the degree of mobility required for each process.

The pre-reaction mechanism will depend on the position and orientation of the molecules in the ice.  This may differ from other experimental setups if a hot-molecule mechanism is at play during deposition.  As mentioned in Section \ref{qual}, it is possible that molecules could be preferentially oriented to react due to energy dissipation following condensation.  If molecules were deposited cold instead of at room temperature, it may result in a smaller fraction of activated reactants.  However, since the two-step model fits for the initial active fraction of reactants, this will not impact our derived reaction barrier.  A hot-molecule mechanism may impact whether a pre-reaction step is required or not, but the barrier should not be impacted since pre-reaction kinetics depend only on molecules that are in the inactive state.

Comparing with previous studies, the pre-reaction step in the case of \citet{Mispelaer2012} is likely only an orientation process, as diffusion is not expected to play a role under their experimental conditions of co-deposition with a great excess of NH$_3$.  We cannot make a comparison to their pre-reaction barrier, however, since they do not observe a temperature dependence for their pre-reaction step, and instead claim that re-orientation is induced by IR photons used to monitor their sample.

In our experiments, is likely that NH$_3$ is the main diffusing species due to its greater mobility over HCOOH.  Thus, the diffusion-limited pre-reaction barrier of $\sim$770K derived in this work can be considered an upper limit for the diffusion barrier for NH$_3$, since it also incorporates orientation. The energy barrier to NH$_3$ diffusing up through water and desorbing has been measured by \citet{Mispelaer2013} to be around 8000K, with values as low as 2000K allowed within the uncertainties.  This range of values is much higher than the value measured in this work.  It is possible that the NH$_3$/HCOOH ice is a more favorable environment for diffusion than a water ice.  Alternatively, it is possible that the presence of charge in the ice impacts diffusion.  For instance, diffusion barriers might be overcome more easily in a charged environment with Coulombic attractions present.  

It is likely that each pre-reaction barrier derived in this work is in reality a distribution of barriers rather than a single value. A pre-reaction barrier distribution would explain the rapid salt growth occurring at low temperatures during the initial warm-up, contrasting with the slow growth observed during isothermal holds (see Figures \ref{codep}, \ref{colay}, and \ref{1mlstep}): the low-energy end of the distribution reacts initially, and at increasing temperatures we are probing the higher-barrier pre-reaction processes. This would also explain the similarity in salt production between the single-temperature experiments and the step experiments (\ref{steps}b-c) since the amount of salt that can form at a given temperature would be controlled by the fraction of pre-reaction barriers that are thermally accessible at that temperature.

Physically, a distribution of energies for the pre-reaction step would not be surprising given that it represents both orientation and diffusion.  These two processes will contribute to a different extent to growth depending on, ice thickness, temperature, and reaction progress.  In particular, the barrier for each process may change over the course of an experiment; for instance, diffusion later in the course of an experiment is likely more difficult as more salt will have built up in between the pure ice layers.
\placefigureEdists

However, it is not obvious why we would observe an Arrhenius-like relationship between pre-reaction rates and temperature in the case of a temperature-dependent distribution of energy barriers. While we see a tail end to the Arrhenius plot at low temperatures in Figure \ref{Arrhenius}, the experiments above $\sim$90K appear linear and thus consistent with Arrhenius behavior.  We performed a toy calculation to evaluate whether a distribution of energy barriers could produce a log-linear relationship between measured rates and temperatures.  We assume that the measured energy barrier increases as a function of temperature since the lower barriers would be overcome at lower temperatures.  Rate constants are calculated using the Arrhenius equation \eqref{arrhenius} with a fixed value of A and with E$_{a}$ varying according to three different distributions: a symmetric equally-spaced distribution from 80\% to 120\%; a symmetric Gaussian distribution from 80\% to 120\%; and an asymmetric equally-spaced distribution from 90\% to 130\%.  

For both linear distributions we see that the rate constants are still linear on the Arrhenius plot despite being derived from different energy barriers.  The Gaussian distribution can appear linear or deviate substantially from linear depending on the choice of standard deviation; we show the results using a moderate value, which could certainly be possible within the scatter of the rate constants we derive for the pre-reaction barrier.  In all three models the derived barrier would be lower than the median barrier, and thus the reported pre-reaction barriers should be considered lower limits.

These plots demonstrate that it is possible under a number of conditions to reproduce linear Arrhenius behavior despite each rate constant being defined by a different energy barrier.  Given the nature of the pre-reaction step, such a distribution is likely to be at play in this system.  It should be noted that, by contrast, the reaction barrier appears to be a single value, independent of temperature and ice environment, as discussed in Section \ref{rxndisc}.

\subsection{Astrophysical Implications}
We next extend our laboratory results to assess the temperature at which the NH$_4^+$/HCOO$^-$ salt formation reaction should be efficient under interstellar conditions.  Because this is an ice process, it should not be impacted by the lower pressure of the interstellar medium compared to the ultra-high vacuum setup.  However, the timescales of astrophysical processes are many orders of magnitude longer than laboratory timescales.  Following \citet{Pontoppidan2008}, we relate the critical temperature $T_{\mathrm{crit}}$ of the reaction under interstellar conditions to that under laboratory conditions:
\begin{equation}
\frac{\tau_{\mathrm{astro}}}{\tau_{\mathrm{lab}}} = \mathrm{exp}\bigg[E_a\bigg(\frac{1}{T_{\mathrm{astro}}} - \frac{1}{T_{\mathrm{lab}}}\bigg)\bigg]
\end{equation}
where $\tau_{\mathrm{astro}}$ and $\tau_{\mathrm{lab}}$ are the e-folding timescales (i.e., a single exponential lifetime) for the reaction in the interstellar medium and the laboratory, respectively, and $T_{\mathrm{astro}}$ and $T_{\mathrm{lab}}$ correspond to the critical temperatures for each environment.  For $\tau_{\mathrm{astro}}$ we assume a typical star formation timescale of 10$^6$ years for a cold cloud, and for $E_a$ we use the reaction barrier of 70K derived in this work.  Since our reaction model is first-order, $\tau_{\mathrm{lab}}$ is simply the inverse of the rate constant (i.e. 1/s$^{-1}$) measured at a given temperature $T_{\mathrm{lab}}$.  Using the rate constants measured in this work we find a critical temperature $T_{\mathrm{astro}}$ of 3K. Such low critical temperature implies that, in the case where reactants are in the appropriate configuration, we expect reaction to proceed under cloud core conditions of $<$10K.  Note that this treatment assumes HCOOH and NH$_3$ are already in the proper position and orientation, i.e. we do not consider pre-reaction steps.  The availability of reactants may therefore limit the degree to which this reaction occurs. 

The impact of pre-reaction barriers on growth kinetics is of particular importance to astrochemical modeling.  Most Monte Carlo models of grain-surface chemistry already account for diffusion using a "hopping" barrier; however, we have demonstrated that other mobility-related barriers may be equally important to ice chemistry.  Orientation has a lower barrier than diffusion, but is not currently incorporated in models and needs to be considered to create a full picture of the chemistry.  The matrix-dependent nature of mobility barriers means that the values of pre-reaction barriers derived in this work are likely not suitable for use in astrochemical models; further work is required to derive appropriate pre-reaction barriers for modeling use.  

Additionally, more work is required to assess the role of ions as initiators of surface chemical pathways.  As discussed by \citet{Woon2011}, surface reactions involving ions offer a promising low-energy pathway to chemical complexity, without the destructive effects of energetic processing.  Woon presents theoretical calculations of several protonated species undergoing barrierless reactions with neutral water molecules; analogous to the gas-phase ion-neutral reactions that are responsible for many gas-phase processes in low-energy interstellar environments, these surface ion-neutral reactions are spontaneous and offer an energetic advantage over neutral-neutral reactions mediated by barriers.  Woon's calculations consider ions as depositing onto water ices, but an in-situ method of ion formation such as that considered in this paper could equally drive these ion-neutral reaction pathways.  As NH$_{3}$ and HCOOH are fairly abundant components of ice mantles, the reaction described in this work has the potential to be a robust source of ion generation, and could increase chemical complexity via low-barrier ion-neutral reactions.  

\section{Conclusion}
The kinetics and mechanisms of the reaction between NH$_3$ and HCOOH were examined by continuously monitoring product formation in layered ices held at constant temperatures.  Based on our results we conclude:

\noindent 1. Under laboratory conditions the reaction between NH$_3$ and HCOOH is accessible as low as 14K and proceeds to completion upon warm-up.

\noindent 2. The proton transfer to form the salt has a barrier of 70K if the molecules are positioned to react.

\noindent 3. A two-step process (a slow pre-reaction step followed by a rapid reaction step) best describes the mechanism of salt growth.  The pre-reaction step is likely a combination between orientation and diffusion of reactant molecules.

\noindent 4. Multilayer ices exhibit a pre-reaction barrier of $\sim$770K while that of single layer ices is only $\sim$400K.  The former includes both diffusion and orientation barriers, while the latter is only orientation.

\noindent 5. The pre-reaction barrier may actually be a distribution of barriers. Pre-reaction barriers at the higher end of the energy distribution are sampled at higher temperatures.  As a result, the values of the pre-reaction barriers listed here are likely upper limits to the actual median barriers.

\noindent 6. NH$_4^+$HCOO$^-$ salt formation in the interstellar medium is potentially possible below 10K provided that the reactants are correctly positioned at neighboring sites.

\bigskip
\noindent J.B.B acknowledges funding from the National Science Foundation Graduate Research Fellowship under Grant DGE1144152.  K.I.\"O. acknowledges funding from the Simons Collaboration on the Origins of Life (SCOL) investigator award.

\section{Appendix 1}
\label{sec_appendix}
Band strengths were derived as follows: a layered ice with NH$_{3}$ beneath HCOOH was heated at 5K/min to 165K. At this point the HCOOH band at 1216 cm$^{-1}$ has disappeared, but HCOOH has not begun to desorb (as monitored by QMS).  This indicates that the reactant has been fully converted, and none is lost to desorption.  Also at this point the salt has not yet crystallized (as seen by IR spectra) or desorbed.  The final peak area for the salt upon reaching 165K is then equated to the inital amount of formic acid.  This method assumes that the salt features do not change as a result of the temperature or the structure of the ice.  This is likely not entirely accurate, and we have assessed the dependencies of each. Figure \ref{BSexp} shows the experiment used to do so. 
\placefigureBSexp

A structure dependence would arise from the fact that the salt forming at low temperatures may not be configured in the same way as salt at higher temperatures.  Salt crystallization occurs above the range of temperatures we are interested in for this study, but more minor bulk rearrangements may occur at lower temperatures.  We explore the magnitude of this effect by comparing the consumption of HCOOH with the formation of salt during warm-up of a layered HCOOH/NH$_3$ ice from 14K to 150K (Figure \ref{BSexp}, left of vertical dotted line). We then assume that there is a one-to-one conversion from HCOOH to HCOO$^-$ and assign a temperature-dependent band strength such that the sum of HCOOH and HCOO$^-$ remains constant throughout the conversion.  When there are low abundances of either species the uncertainties of the derived band strengths are large, but we focus on the range 60-130K in which most experiments are performed.  Here, the determined band strength varies by less than 5\%. 

We next assess the temperature dependence of the salt band strength for an ice with a constant structure by comparing the strength of the salt feature at 165K with the strength of the same feature after the salt has been cooled down (Figure \ref{BSexp}, right of vertical dotted line).  After reaching 165K the salt was cooled back to 11K at 5K/min and then heated once again at 5K/min until fully desorbed.  The band strength does change by $\sim$5\% over this temperature range, and the change in band strength is fully reversible. 

We do not attempt to correct for these dependencies in performing our fitting, as this would introduce many more uncertainties but not ultimately change the values substantially.  Instead, these uncertainties are incorporated into the error analysis.

\bibliographystyle{apj}

\end{document}